\documentclass[aps]{revtex4-1}
\usepackage{graphicx}
\usepackage{amsmath}
\usepackage{makeidx}
\usepackage{amsfonts}
\usepackage{amssymb}
\usepackage{tabularx}
\usepackage[dvipsnames]{xcolor}

\begin{document}

\title{Statistical physics of principal minors: Cavity approach}

\author{A. Ramezanpour}
\email{aramezanpour@gmail.com}
\affiliation{Department of Physics, College of Sciences, Shiraz University, Shiraz 71454, Iran}
\affiliation{Medical Systems Biophysics and Bioengineering, Leiden Academic Centre for Drug Research, Faculty of Science, Leiden University, Leiden, The Netherlands}	

\author{M. A. Rajabpour}
\email{mohammadali.rajabpour@gmail.com}
\affiliation{Instituto de Fisica, Universidade Federal Fluminense, Av. Gal. Milton Tavares de Souza, s/n, Campus da Praia Vermelha Sao Domingos, 24210-346 Niteroi-RJ, Brasil}

\date{\today}

\begin{abstract}
Determinants are useful to represent the state of an interacting system of (effectively) repulsive and independent elements, like fermions in a quantum system and training samples in a learning problem. A computationally challenging problem is to compute the sum of powers of principal minors of a matrix which is relevant to the study of critical behaviors in quantum fermionic systems and finding a subset of maximally informative training data for a learning algorithm.     
Specifically, principal minors of positive square matrices can be considered as statistical weights of a random point process on the set of the matrix indices. The probability of each subset of the indices is in general proportional to a positive power of the determinant of the associated sub-matrix. We use Gaussian representation of the determinants for symmetric and positive matrices to estimate the partition function (or free energy) and the entropy of principal minors within the Bethe approximation. The results are expected to be asymptotically exact for diagonally dominant matrices with locally tree-like structures. We consider the Laplacian matrix of random regular graphs of degree $K=2,3,4$ and exactly characterize the structure of the relevant minors in a mean-field model of such matrices. No (finite-temperature) phase transition is observed in this class of diagonally dominant matrices by increasing the positive power of the principal minors, which here plays the role of an inverse temperature. 
\end{abstract}


\maketitle

\section{Introduction}\label{S0}
A principal minor of a matrix $\mathbf{A}$ is determinant of a square submatrix which is formed by a number of rows and columns with the same indices from the matrix. This provides a measure of independence for the selected subset of rows (or columns) which is useful for instance in sampling problems where diversity matters or in computing the entropy of physical systems with fermionic statistics.    
The sum of powers of principal minors (SPPM) of a matrix appears in various areas of physics and mathematics \cite{morteza-2023}. SPPM finds applications in the study of determinantal point processes \cite{Kulesza2012,Gillenwater2014}, R\'enyi entropy of free fermions and quantum spin chains \cite{NR2016}, and partition function of the Hubbard model \cite{morteza-2023}. When the power is one, the sum is a simple determinant that can be computed in polynomial time. However, for other powers, the problem is generally considered to be NP-hard, indicating that there may not be a polynomial time algorithm to calculate them \cite{Gurvits2005,Ohsaka2021}. Nonetheless, recent advancements have been made in the approximate analytical computation of these quantities \cite{morteza-2023}. In this paper we study statistical properties of the principal minors of positive, symmetric, and diagonally dominant matrices which can be represented by Gaussian integrals and estimated by the cavity method of statistical physics. We employ this method to characterize the spectral entropy of principal minors in this subclass of matrices and utilize it as an approximate optimization algorithm to find the maximal minor configurations.    

The principal minors of a matrix has a very natural meaning in the context of the graph theory. Consider, the adjacency matrix of a graph $G$ with $n$ nodes, which is a matrix that describes the connections between the nodes of the graph, where the entry in row $i$ and column $j$ of the matrix is $1$ if there is an edge from node $i$ to node $j$, and $0$ otherwise. Then the Laplacian matrix of a graph is defined as the difference between the degree matrix and the adjacency matrix of the graph. The degree matrix is a diagonal matrix where the entry in the $i$-th row and $i$-th column is equal to the degree of the $i$-th node, i.e., the number of edges incident to it, and the non-diagonal entries are $0$. The Laplacian matrix $\mathbf{L}$ is a symmetric, positive semi-definite matrix which means all the principal minors are non-negative. It has numerous applications in graph theory notably in the enumeration of the spanning trees. A spanning tree of an undirected graph is a tree that includes all of the graph's vertices, while using only a subset of its edges, such that no cycles (closed paths) are formed.
The matrix-tree theorem, see for example \cite{BondyMorty1976},  then states that the number of spanning trees of the graph $G$ is equal to any cofactor of the Laplacian matrix $\mathbf{L}$. Specifically, if we remove any row and column (say $i$) from $\mathbf{L}$ to obtain a $(n-1)\times(n-1)$ matrix $\mathbf{L}[i]$, then the number of spanning trees of $G$ is given by the determinant of $\mathbf{L}[i]$.

There is a generalization of matrix-tree theorem which is called principal minor matrix tree theorem which is used to enumerate the number of rooted spanning forests \cite{Chaiken1982}. A forest is formed by a set of disjoint tree subgraphs of the graph $G$. A rooted spanning forest is a subgraph of $G$ that is a forest and contains all vertices of the graph. And each tree in the forest is rooted at a designated root vertex in that tree. Let $\mathbf{L}_{\mathbf{c}}$ be the submatrix of $\mathbf{L}$ formed by selecting the rows and columns corresponding to the vertices in $\mathbf{c}$ then the number of spanning forests rooted at $\mathbf{c}$ is $\mathrm{det} \mathbf{L}_{\mathbf{c}}$.
This interpretation of the principal minors makes the definition of the weighted partition function of the rooted spanning forests as the sum of powers of principal minors natural. In this paper we interpret the power as the inverse of a temperature-like parameter. In the limit of large powers, i.e. small temperatures, the SPPM gives the principal minor with the largest contribution which is a desired quantity in determinantal point processes \cite{Kulesza2012} and the study of random spanning trees \cite{Anari2022}.

Here we introduce a novel statistical model in which the principal minors of a positive semi-definite matrix serve as the statistical weights. Although the matrix need not necessarily be a graph Laplacian, our focus is primarily on this type of matrix. To address this problem, we utilize the cavity method (Bethe approximation or belief propagation)\cite{marc-epjb-2001,marc-jstat-2003,bp-2003,marc-book-2009}, which is commonly used in spin glasses but has not been previously applied to studying the sum of powers of principal minors. Examples of applications of the method in similar problems, mainly in the study of spectral properties of sparse matrices, can be found in Refs. \cite{semerjian-jphysa-2002,rogers-pre-2008,huang-arxiv-2009,rogers-pre-2009,metz-pre-2010,kabashima-jphys-2010,biroli-ptheory-2010}. The cavity method should be asymptotically exact in interacting systems which have a locally tree-like interaction graph \cite{marc-book-2009}. The sum of principal minors to power $\beta$ here is considered as the partition function of principal minors of the matrix at inverse temperature $\beta$.  We employ the Gaussian representation of determinants to find estimations of the free energy and entropy of the relevant principal minors for Laplacian of random regular graphs of degree $K=2,3,4$. These results along with an exact study of diagonally dominant matrices defined on fully connected graphs indicate on the absence of a finite-temperature phase transition in this class of matrices. In a diagonally dominant matrix, the magnitude of a diagonal element $|A_{ii}|$ in each row is larger than or equal to the sum of magnitudes of the off-diagonal elements $\sum_{j\neq i}|A_{ij}|$. 

The nontrivial structure of the ground states in random regular graphs of degree $K=2$ (or chains) however results in numerical instabilities for large $\beta$ and degrades the efficiency of the simulated annealing algorithm in finding a maximal principal minor in these systems; the entropy density of relevant minor configurations is discontinuous at zero temperature, jumping to a nonzero value for any finite temperature. The situation is much better for (random) Laplacian of graphs with larger degrees $K>2$ as one approaches the mean-field limit, where the ground states have a trivial structure in the configuration space of the principal minors. Nevertheless, it is possible to observe discontinuities in the entropy density at zero temperature by introducing a chemical potential which controls the number of present indices in the minors. These zero-temperature transitions are similar in nature to those of the smaller degree $K=2$. We also use the zero-temperature limit of the Bethe equations, the so called MaxSum equations \cite{marc-book-2009,baldassi-polito-2009}, as an approximate optimization algorithm to find a maximal principal minor of diagonally dominant matrices. The standard algorithms are in general based on the spectral decomposition of the matrix with a time complexity of order $N^3$ for an arbitrary matrix of dimension $N$ \cite{Kulesza2012}. The computational complexity can of course be reduced to $N$ for sampling of principal minors of dimensions $n\ll N$ \cite{Anari2022,valko-advni-2019,poulson-phil-2020}. The time complexity of the MaxSum algorithm here is proportional to $N$ for sparse and diagonally dominant matrices where the (Gaussian) Bethe equations are expected to work. 

The paper is structured as follows. In Sec. \ref{S1} we define the main quantities of the problem. In Sec. \ref{S2} we present the results obtained by the Gaussian representation of the determinants. This section includes subsections that deal with the finite- and zero-temperature limits of the Bethe equations. The summary of results and concluding remarks are given in Sec. \ref{S3}. In the appendices, we give a brief introduction to the Bethe approximation (\ref{app:bp}), describe the details of the population dynamics which is used to solve the Bethe equations (\ref{app:pop}), write the simplified Bethe equations for random regular graphs (\ref{app:rrg}), and present an exact treatment of principal minors in matrices defined by homogeneous fully-connected graphs (\ref{app:MF}).

\section{The problem statement and definitions}\label{S1}
Consider a positive square matrix $\mathbf{A}$ of size $N$ with elements $A_{ij}$ indexed by $i,j=1,\dots,N$. The nonzero elements of $\mathbf{A}$ define the structure or interaction graph of the matrix. The set of neighbors of node $i$ in this graph is denoted by $\partial i$. The $2^N$ principal minors $\mathrm{det}\mathbf{A}(\mathbf{c})$ are labeled with configurations $\mathbf{c}=\{c_i=0,1:i=1,\dots,N\}$. That is $c_i=1$ shows that row (column) $i$ is included in the square sub-matrix $\mathbf{A}(\mathbf{c})$.
Each configuration is assigned a Boltzmann weight 
\begin{align}
P_{\beta}(\mathbf{c})=\frac{[\mathrm{det}\mathbf{A}(\mathbf{c})]^{\beta}}{Z(\beta)}=\frac{e^{-\beta E(\mathbf{c})}}{Z(\beta)},
\end{align}
where the associated energy function reads
\begin{align}
E(\mathbf{c})=-\ln[\mathrm{det}\mathbf{A}(\mathbf{c})],
\end{align}
and the positive power $\beta$ plays the role of an inverse temperature. The energy of all-zero minor configuration is zero that is $\mathrm{det}\mathbf{A}(\mathbf{0})=1$. 
Here the partition function (normalization factor) is
\begin{align}\label{ZP}
Z(\beta)=\sum_{\mathbf{c}} [\mathrm{det}\mathbf{A}(\mathbf{c})]^{\beta}=\int de e^{-\beta N(e-\frac{s(e)}{\beta})}=e^{-\beta Nf(\beta)},
\end{align}
with the energy density $e=E/N$ and the free energy density $f(\beta)=-[\ln Z(\beta)]/(\beta N)$. 
The entropy density is defined by $s(e)=\ln(\Omega(e))/N$ where $\Omega(e)$ is the number of configurations with energy density $e$.  
In the thermodynamic limit $N\to \infty$, the integrand is concentrated on the mean energy, that is $f=e-\frac{1}{\beta}s$.

In the following, we are going to change the parameter $\beta$ to study the energy and entropy landscape of the SPPM problem given in Eq. \ref{ZP}. The range of energy values (principal minors of $\mathbf{A}$) represent the relevant minors for different values of $\beta$. Specially, the minimum energy value and configuration, i.e., the ground state(s) are obtained at zero temperature ($\beta \to \infty$). The entropy spectrum $s(e)$ shows the number of such relevant sub-matrix configurations and the free energy $f$ is a measure of the sum over the weights of these configurations at inverse temperature $\beta$.

From the above system we can compute the R\'enyi entropy of the probability distribution $P_{\beta}(\mathbf{c})$, 
\begin{align}
R_{\beta}(n)=\frac{1}{1-n}\ln\left(\sum_{\mathbf{c}}P_{\beta}^n(\mathbf{c})\right)=\frac{1}{1-n}[\ln Z(n\beta)-n \ln Z(\beta)].
\end{align}
Specifically
\begin{align}
R_{\beta}(1)=-\sum_{\mathbf{c}}P_{\beta}(\mathbf{c})\ln P_{\beta}(\mathbf{c}),
\end{align}
which can be obtained by a Legendre transformation of the free energy
\begin{align}
\frac{R_{\beta}(1)}{N}=s(\beta)=\beta(e(\beta)-f(\beta)).
\end{align}
Note that here $e(\beta)$ is the average energy density at inverse temperature $\beta$. A parametric plot of the above quantity gives the entropy $s(e)$ as a function of energy. Here one is interested in the values of the free energy, entropy, and energy of the relevant minors at different inverse temperatures $\beta$, the structure of these minors in the configuration space and the nature of possible phase transitions in this system.

\section{Gaussian representation}\label{S2}
Let us assume that $\mathbf{A}$ is a positive and symmetric matrix. Thus all the square sub-matrices are positive and symmetric.
Given a configuration $\mathbf{c}$ of the indices we define the sub-matrix $\mathbf{A}(\mathbf{c})$ with elements
\begin{align}
A_{ij}(\mathbf{c})=c_iA_{ij}c_j+(1-c_i)\delta_{i,j}.
\end{align}
A diagonal regularization takes care of the case $i=j$ and $c_i=0$.
Then, we employ the Gaussian integrals to write
\begin{align}
\mathrm{det}\mathbf{A}(\mathbf{c})=\frac{(2\pi)^N}{\left(\int_{-\infty}^{+\infty} \prod_{i=1}^N dx_i e^{-\frac{1}{2}\sum_{i,j}x_i(c_iA_{ij}c_j+(1-c_i)\delta_{i,j})x_j}\right)^2}.
\end{align}
In this way
\begin{align}
\ln \mathrm{det}A(\mathbf{c})=N \ln (2\pi)-2\ln \left(\int \prod_i dx_i e^{-\frac{1}{2}\sum_{i,j}x_i(c_iA_{ij}c_j+(1-c_i)\delta_{i,j})x_j} \right),
\end{align}
or
\begin{align}
\ln \mathrm{det}A(\mathbf{c})=N\ln (2\pi)-2Ng(\mathbf{c}),
\end{align}
where for brevity we defined
\begin{align}\label{gc}
\int \prod_i dx_i e^{-\frac{1}{2}\sum_{i,j}x_i(c_iA_{ij}c_j+(1-c_i)\epsilon\delta_{i,j})x_j}=e^{Ng(\mathbf{c})}.
\end{align}
Recall that $E(\mathbf{c})= -\ln \mathrm{det}\mathbf{A}(\mathbf{c})$, therefore,
\begin{align}
Z(\beta)=\sum_{\mathbf{c}} e^{-\beta E(\mathbf{c})}=e^{\beta N\ln (2\pi)}\sum_{\mathbf{c}} e^{-2\beta Ng(\mathbf{c}) }.
\end{align}
Note that $g(\mathbf{c})$ is a global function of the $c_i$. In the following, we write this quantity as a sum of local energy contributions by introducing other auxiliary variables to the problem. This allows us to apply the standard methods of estimating the free energy of systems with a local energy function.

\subsection{Bethe approximation of the Gaussian integral}\label{S21}

In this section we use Bethe approximation to find an estimation of the Gaussian integral in Eq. \ref{gc}. In Appendix \ref{app:bp}, we briefly explain the method for a simple statistical model, see also Refs. \cite{marc-epjb-2001,marc-jstat-2003,bp-2003,marc-book-2009,baldassi-polito-2009}. The Bethe or belief propagation (BP) equations for the Gaussian integrals are recursive equations for cavity probability distributions $m_{i\to j}(x_i)$. This is probability density of $x_i$ in the absence of interaction with variable $x_j$ assuming that the interaction graph defined by $\mathbf{A}(\mathbf{c})$ has a tree structure. To write these equations, we consider the local interactions of $x_i$ with its neighboring variables which are assumed to be statistically independent of each other,
\begin{align}
m_{i\to j}(x_i) \propto e^{-\frac{1}{2}((1-c_i)+c_iA_{ii}) x_i^2}\prod_{k\ne i,j} \left(\int dx_k e^{-x_ic_iA_{ik}c_kx_k}m_{k\to i}(x_k) \right).
\end{align}
It is known that even in interaction graphs which are not tree (loopy graphs) these equations converge to a unique solution as long as $\mathbf{A}(\mathbf{c})$ is a diagonally dominant matrix \cite{GBP1-ieee-2008,GBP2-ieee-2014}, that is $|A_{ii}|\ge \sum_{j\neq i}|A_{ij}|$. 

Now consider the following ansatz for the BP messages
\begin{align}
m_{i\to j}(x_i) \propto e^{-\frac{x_i^2}{2v_{i\to j}}}.
\end{align}
Note that because of the symmetry of the problem  we are assuming that $\langle x_i\rangle=0$. This results in a set of BP equations for the variances,
\begin{align}\label{vij}
\frac{1}{v_{i\to j}}=(1-c_i)+c_iA_{ii}-\sum_{k\ne i,j} c_iA_{ik}^2c_k v_{k\to i}.
\end{align}
That is, the Gaussian ansatz is consistent with the structure of the Gaussian BP equations.  

Given the solution to the BP equations, we write $g(\mathbf{c})$ in terms of the local contributions of the variables and interactions to the Gaussian integral \cite{baldassi-polito-2009},
\begin{align}
Ng(\mathbf{c})=\sum_i \Delta g_i(\mathbf{c})-\sum_{i<j} \Delta g_{ij}(\mathbf{c}),
\end{align}
where
\begin{align}
e^{\Delta g_i} &=\int dx_i e^{-\frac{1}{2}((1-c_i)+c_iA_{ii}) x_i^2}\prod_{j\ne i} \left(\int dx_j e^{-x_ic_iA_{ij}c_jx_j}m_{j\to i}(x_j)\right),\\
e^{\Delta g_{ij}} &=\int dx_idx_j  e^{-x_ic_iA_{ij}c_jx_j}m_{i\to j}(x_i)m_{j\to i}(x_j).
\end{align}
Later, we will also need a cavity contribution of the variables $\Delta g_{i\to j}$, which is like $\Delta g_i$ but excluding one of the neighboring interactions,
\begin{align}
e^{\Delta g_{i\to j}}=\int dx_i e^{-\frac{1}{2}((1-c_i)+c_iA_{ii}) x_i^2}\prod_{k\ne i,j} \left(\int dx_k e^{-x_ic_iA_{ik}c_kx_k}m_{k\to i}(x_k) \right).
\end{align}

Within the Gaussian ansatz for the BP messages, the above expressions for the local variable and interaction contributions are simplified to 
\begin{align}\label{Dg}
2\Delta g_i &=\ln(2\pi v_i),\\
2\Delta g_{ij} &=-\ln\left(v_{i\to j}v_{j\to i}\mathrm{det}(B(ij))\right),
\end{align}
where
\begin{align}\label{Bij}
\frac{1}{v_i} &=(1-c_i)+c_iA_{ii}-\sum_{k\ne i} c_iA_{ik}^2c_k v_{k\to i},\\
B(ij) &=\begin{pmatrix} \frac{1}{v_{i\to j}} & c_iA_{ij}c_j \\ c_iA_{ij}c_j & \frac{1}{v_{j\to i}} \end{pmatrix}.
\end{align}
Note that the sign of off-diagonal elements $A_{ij}$ is irrelevant as long as the Gaussian Bethe equations are valid. That is, we can change the sign of any such element of the matrix without changing the determinant or the energy of the system within the above approximation.

In the next subsection, we use the above expressions for the determinants of $\mathbf{A}(\mathbf{c})$ to compute the partition function of the original problem.

\subsection{A higher level Bethe approximation of the partition function}\label{S22}
Now we are ready to do the sum over the configurations  
\begin{align}
\sum_{\mathbf{c}} e^{-2\beta Ng(\mathbf{c}) }=e^{-\beta N f_g},
\end{align}
which is needed in the partition function
\begin{align}
Z(\beta)=e^{-\beta N(f_g-\ln (2\pi))}.
\end{align}
Here we defined $f_g$ as the nontrivial contribution to the free energy $f=f_g-\ln (2\pi)$.

In the previous subsection we wrote $g(\mathbf{c})$ in terms of the $\Delta g_i$ and $\Delta g_{ij}$ which in turn depend on the BP messages $v_{i\to j}$. These messages satisfy the BP equations with a unique solution for diagonally-dominant matrices. Therefore, we can write 
\begin{align}\label{Zg}
\sum_{\mathbf{c}} e^{-2\beta Ng(\mathbf{c}) }=\sum_{\mathbf{c}}\sum_{\mathbf{v}_{\to}} e^{-2\beta (\sum_i \Delta g_i-\sum_{i<j}\Delta g_{ij}) } \mathbb{I}(\mathbf{v}_{\to}),
\end{align}
that is considering the $v_{i\to j}$ as auxiliary variables which are constrained by the indicator function $\mathbb{I}(\mathbf{v}_{\to})=\prod_{i\to j}\delta(v_{i\to j}-v_{i\to j}^{BP})$ to satisfy the BP equations.

Again we can resort to the Bethe approximation to estimate the sum over the extended set of variables $\mathbf{c},\mathbf{v}_{\to}$. We can do this because both the $g(\mathbf{c})$ and the BP constraints on the $v_{i\to j}$ are local functions of these variables. If the graph associated to matrix $\mathbf{A}$ has a tree structure then the interaction graph of the $c_i$ and the auxiliary variables $v_{i\to j}$ is also a tree. Consider the cavity probability distribution $M_{i\to j}(c_i,v_{i\to j})$ of the variables in the absence of interactions with node $j$. The cavity variables $\{(c_k,v_{k\to i}): k\in \partial i\setminus j\}$ are independent of each other in a tree interaction graph. Thus $M_{i\to j}(c_i,v_{i\to j})$ is proportional to the product of the cavity probabilities $\{M_{k\to i}(c_k,v_{k\to i}): k\in \partial i\setminus j\}$ for variables that are consistent with the hard constraint $\mathbb{I}(v_{i\to j})$. In addition, $M_{i\to j}(c_i,v_{i\to j})$ is also proportional to the Boltzmann factor $e^{\Delta g_i-\Delta g_{ij}}=e^{\Delta g_{i\to j}}$ which gives the statistical weight of the cavity variables (see Eq. \ref{Zg}). Therefore, the higher-level BP equations for the cavity probability distributions read
\begin{align}\label{BPM}
M_{i\to j}(c_i,v_{i\to j}) \propto \prod_{k\in \partial i\setminus j}\left( \sum_{c_k}\int dv_{k\to i} M_{k\to i}(c_k,v_{k\to i}) \right)
\times \mathbb{I}(v_{i\to j})e^{-2\beta \Delta g_{i\to j}}.
\end{align}
We solve these equations by a population dynamics algorithm which is explained in Appendix \ref{app:pop}. The probability distributions $M_{i\to j}(c_i,v_{i\to j})$ are represented by populations of the variables $P_{i\to j}=\{(c_i^a,v_{i\to j}^a):a=1,\dots,\mathcal{N}_p\}$ on each directed link $(i\to j)$. Then, members of the populations from the right hand side of the equation are selected to update the members of the population in left hand side according to the Boltzmann weights of the variables and the hard constraints of the BP equations \cite{lenka-thesis}. All members of the populations are updated in a single iteration of the algorithm. In the stationary state of the population dynamics, say after $t_{eq}$ iterations, we obtain an estimation of the free energy
\begin{align}
Nf_g=\sum_i \Delta f_i-\sum_{i<j} \Delta f_{ij},
\end{align}
with local free energies that are given by
\begin{align}\label{BPF}
e^{-\beta \Delta f_i} &=\sum_{c_i} \prod_{j\in \partial i}\left( \sum_{c_j}\int dv_{j\to i} M_{j\to i}(c_j,v_{j\to i}) \right) e^{-2\beta \Delta g_i},\\
e^{-\beta \Delta f_{ij}} &=\sum_{c_i,c_j} \int dv_{ij} e^{-2\beta \Delta g_{ij}} M_{i\to j}(c_i,v_{i\to j}) M_{j\to i}(c_j,v_{j\to i}). 
\end{align}
Here $v_{ij}=\{v_{i\to j},v_{j\to i}\}$.

\begin{figure}
\includegraphics[width=16cm]{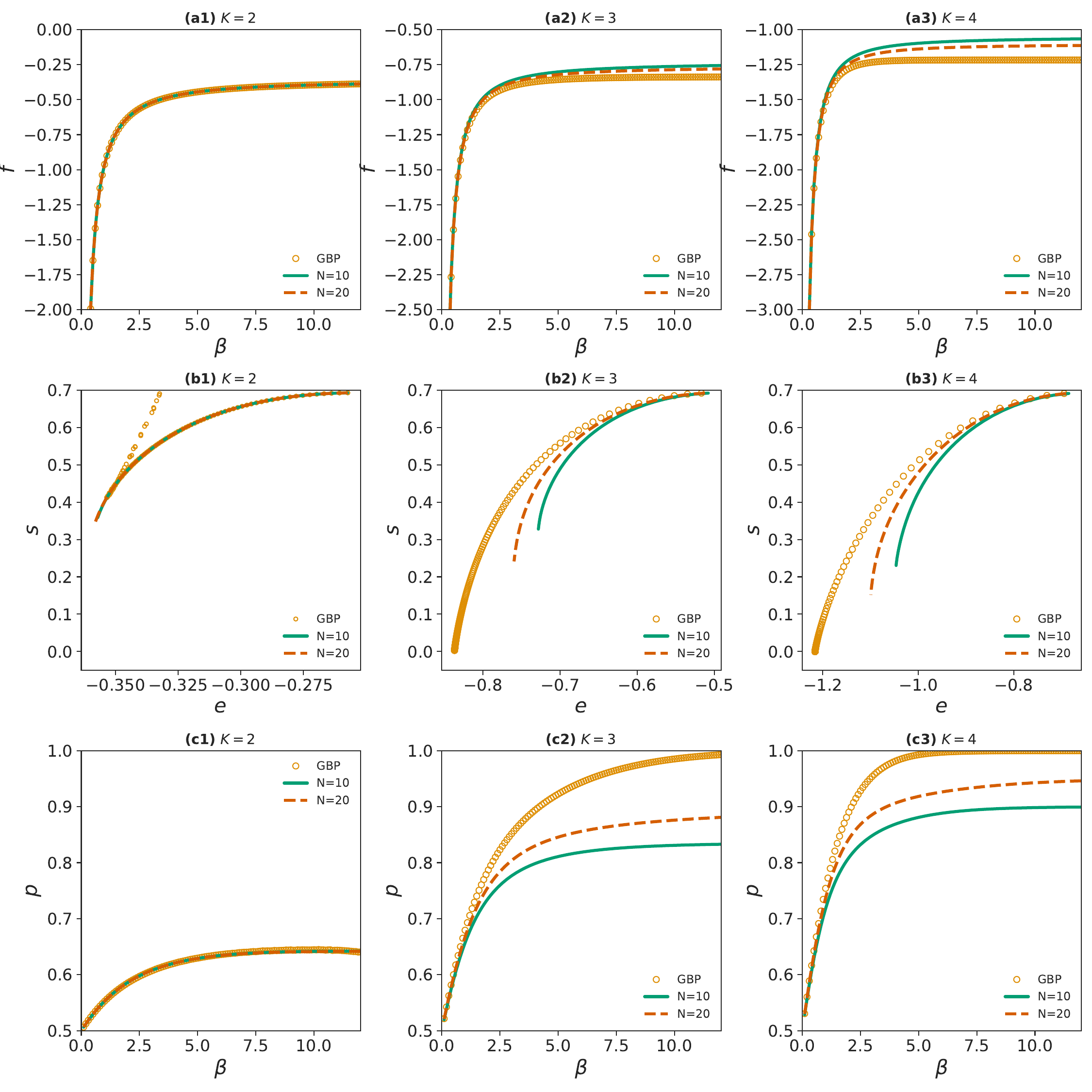} 
\caption{The asymptotic behavior of minors for Laplacian of random regular graphs of degree $K=2,3,4$. The free energy $f$, entropy vs energy $s(e)$, and probability of presence $p$ are reported (Gaussian BP) and compared with exact numerical results for small sizes ($N=10,20$). $\beta$ is the inverse of temperature in the partition function of the principal minors. The parameters of the population dynamics algorithm are: population size $\mathcal{N}_p=10^4$, equilibration time $t_{eq}=10^6$, and averaging time $\Delta t_{avg}=10^4$.}\label{fig1}
\end{figure}

The above computations are simplified for homogeneous interaction graphs where by symmetry all equations for directed links $(i\to j)$ reduce to a single equation. Here we consider the Laplacian of random regular graphs (RRGs) with the same degree (number of neighbors) $K$ for all nodes. See table \ref{tab1} for definitions of the matrices we shall study in this work. In Appendix \ref{app:rrg} we present the resulted equations which again are solved by a population dynamics algorithm. In this case a single population of the messages $(c_i,v_{i\to j})$ is enough to solve the higher-level BP equations. Figure \ref{fig1} shows the main quantities computed in this way for random regular graphs in the thermodynamic limit ($N\to \infty$). The results are obtained by a population of size $\mathcal{N}_p=10^4$ after $t_{eq}=10^6$ iterations of the population dynamics to reach a steady state where the average quantities are stationary.
For comparison we also present the exact numerical results for small problem instances ($N=10,20$). As the figure shows finite size effects are more pronounced for larger degrees $K=3,4$, where the entropy $s(e)$ decreases continuously to zero by increasing $\beta$. And the presence probability $p=\langle c_i\rangle$ approaches  monotonically to $1$ for large $\beta$, where nearly all the indices are present in the relevant minors. No sign of a finite-temperature phase transition is observed here for $K=3,4$. By increasing $\beta$, the Boltzmann weight is smoothly concentrated more and more on minor configurations with a larger number of present indices.

\begin{table}[ht]
\begin{tabularx}{\textwidth}{|X|X|}
\hline
{\bf Matrix} & {\bf Elements} \\
\hline
Laplacian of RRGs (Sec. \ref{S22}) & $A_{ii}=K, A_{ij}=-1 (j \in \partial i), |\partial i|=K$ \\
\hline
quasi-Laplacian of RRGs (Sec. \ref{S22}) & $A_{ii}=\Lambda, A_{ij}=-\gamma/K (j \in \partial i), |\partial i|=K, 0<\gamma\le \Lambda$ \\
\hline
random Laplacian of RRGs (Sec. \ref{S23}) &  $A_{ii}=-\sum_{j\ne i}A_{ij}, A_{ij}\in (-2,0) (j \in \partial i), |\partial i|=K$ \\
\hline
scaling limit I of complete graphs (App. \ref{app:MF}) & $A_{ii}=N-1, A_{ij}=-1 (j \ne i)$ \\
\hline
scaling limit II of complete graphs (App. \ref{app:MF}) & $A_{ii}=\Lambda, A_{ij}=-\gamma/N (j \ne i, 0<\gamma\le \Lambda)$ \\
\hline
\end{tabularx}
\caption{The matrices studied in this paper with a brief description of the defining parameters. These are $N\times N$ matrices with indices $i,j=1,\dots,N$. The random regular graphs (RRGs) have degree $K$ and $\partial i$ denotes the set of neighbors of $i$. In random matrices the off-diagonal elements are independent random numbers uniformly distributed in $(-2,0)$.}\label{tab1}
\end{table}

\begin{figure}
\includegraphics[width=12cm]{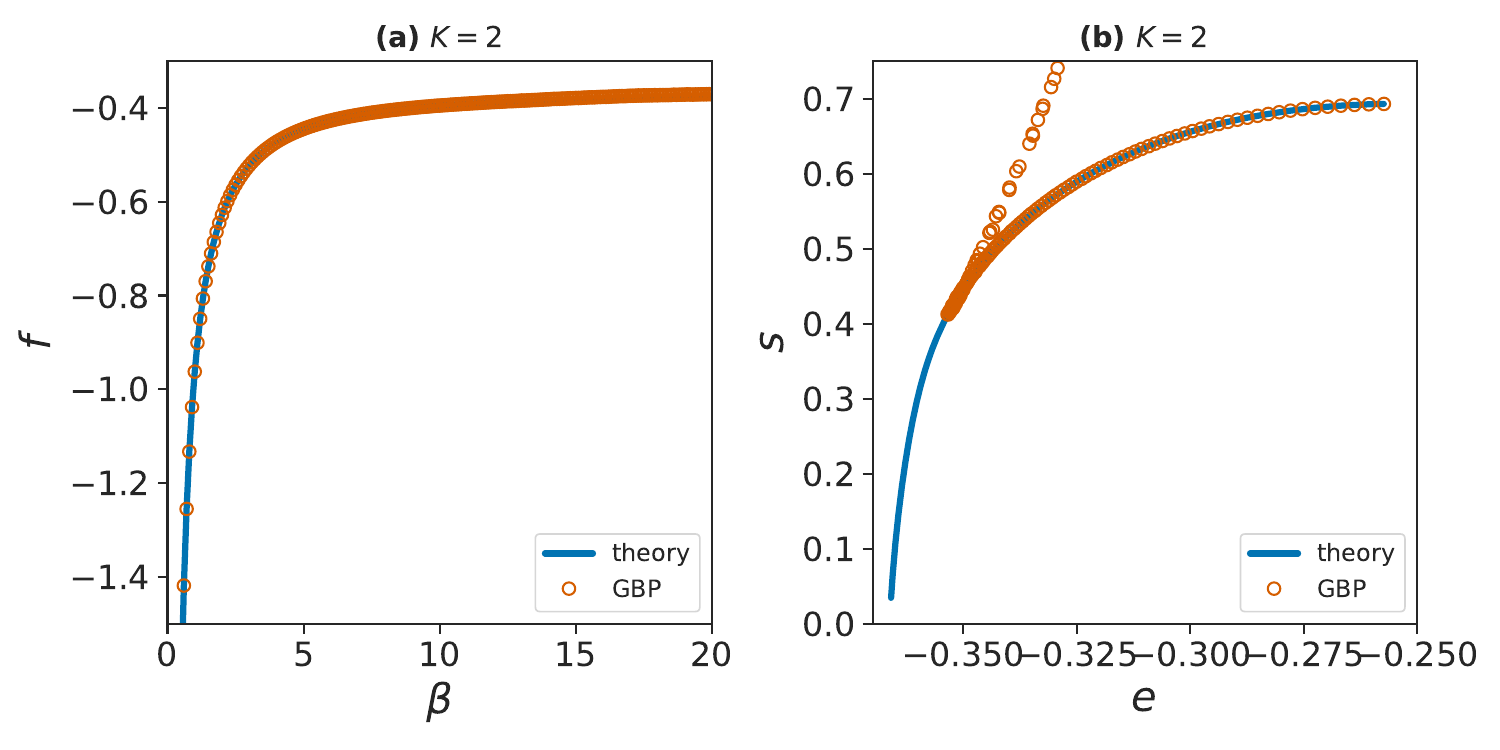} 
\caption{The asymptotic behavior of minors for Laplacian of a closed chain. The free energy $f$ and entropy vs energy $s(e)$ are reported (Gaussian BP) and compared with the exact solution (theory) of Ref.\cite{morteza-2023}. The parameters of the population dynamics algorithm are: population size $\mathcal{N}_p=10^4$, equilibration time $t_{eq}=10^6$, and averaging time $\Delta t_{avg}=10^4$.}\label{fig2}
\end{figure}

The case $K=2$ displays more interesting behaviors with a discontinuity in the entropy for large $\beta$ and a probability $p$ which approaches to a nontrivial value as $\beta$ goes to infinity. We know that there are a sub-exponential number of ground states with extensive Hamming distances in the configuration space \cite{morteza-2023}. The ground states are dimer coverings with pairs of adjacent nodes which are separated with one unmatched node. However, no finite-temperature phase transition is expected to happen also for $K=2$ because the entropic contributions to the free energy can easily destroy the ordered states of this one-dimensional system. We also observe an instability in the BP equations for large $\beta$ close to the discontinuity. Figure \ref{fig2} compares the exact theoretical solution of Ref. \cite{morteza-2023} with the result we obtain by the above Gaussian BP equations. This clearly shows the region of instability where the entropy $s(e)$ is not anymore a concave function of the energy density. In fact, the number of iterations $t_{eq}$ which is necessary to get close to the exact theoretical solution increases rapidly as one approaches the point of discontinuity.

To have a better picture of the phase space of these problems we consider two coupled replicas of the system at equilibrium. By controlling the strength of coupling between the replicas we can investigate the structure of the relevant minors around a given point of the configuration space. We define the partition function of two replicas as
\begin{align}
Z(\beta,h)=\sum_{\mathbf{c},\mathbf{c}'}e^{-\beta[E(\mathbf{c})+E(\mathbf{c}')-h\sum_i\delta_{c_i,c_i'}]}.
\end{align}
This gives the free energy $f$ of the replicas as a function of $\beta$ and the coupling $h$. In terms of the energy densities and the overlap $q=\frac{1}{N}\sum_i \delta_{c_i,c_i'}$, we have
\begin{align}
Z(\beta,h)=e^{-\beta N f}=\int de de' dq e^{-\beta N[e+e'-h q-\frac{1}{\beta}s(e,e':q)]}.
\end{align}
Here the entropy density is $s(e,e':q)=\frac{1}{N}\ln\Omega(e,e':q)$ where $\Omega(e,e':q)$ is the number of two-replica configurations of energy densities $e$ and $e'$ with overlap $q$. In the thermodynamic limit, by the saddle point approximation one obtains
\begin{align}
f &= 2e-hq-\frac{1}{\beta}s(e:q),\\
s(e:q) &=\beta(2e-hq-f),
\end{align}
where the last equation for $s(e:q)$ is obtained after a Legendre transformation. Note that by symmetry the equilibrium energy densities of the two replicas are the same. 

\begin{figure}
\includegraphics[width=16cm]{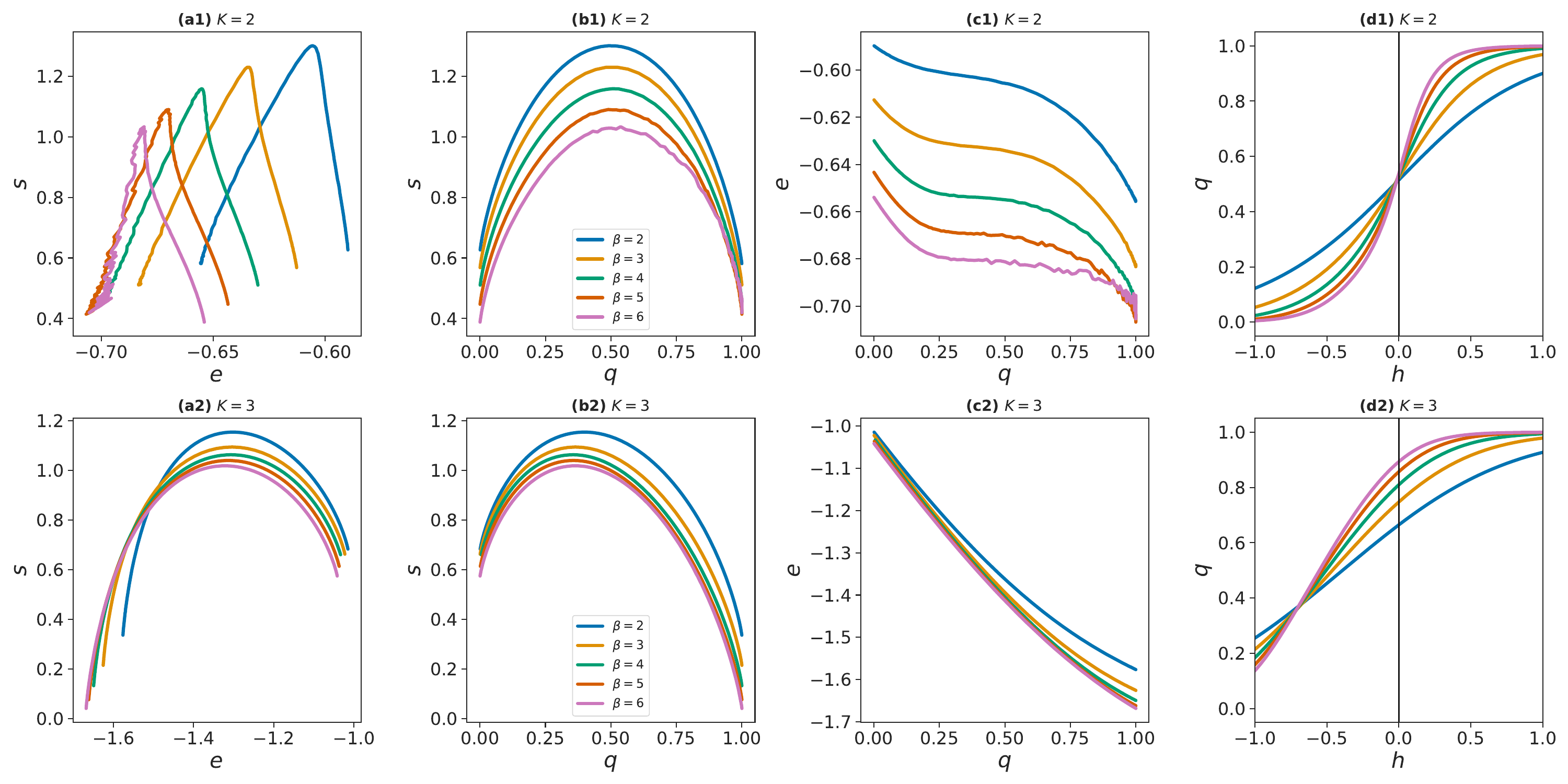} 
\caption{The asymptotic behavior of two coupled minor configurations (replicas) for Laplacian of random regular graphs of degree $K=2,3$. $h$ is the strength of coupling and $q$ is the overlap of the two replicas. The parameters of the population dynamics algorithm are: population size $\mathcal{N}_p=10^4$, equilibration time $t_{eq}=10^6$, and averaging time $\Delta t_{avg}=10^4$.}\label{fig3}
\end{figure}

The higher-level BP equations here are marginal probability distributions for the replica variables
\begin{multline}
M_{i\to j}(c_i,c_i',v_{i\to j},v_{i\to j}') \propto e^{\beta h\delta_{c_i,c_i'}}\prod_{k\in \partial i\setminus j}\left( \sum_{c_k,c_k'}\int dv_{k\to i}dv_{k\to i}' M_{k\to i}(c_k,c_k',v_{k\to i},v_{k\to i}') \right)\\
\times \mathbb{I}(v_{i\to j})\mathbb{I}(v_{i\to j}')e^{-2\beta (\Delta g_{i\to j}+\Delta g_{i\to j}')}.
\end{multline}
As before we consider random regular graphs and solve these equations by population dynamics. The algorithm is very similar to
the one presented in Appendix \ref{app:pop} except that we have the Boltzmann weight $e^{\beta h\delta_{c_i,c_i'}}$ that couples the two replicas. Figure \ref{fig3} shows how the entropy, energy, and overlap of the two systems are related to each other in random regular graphs of degree $K=2,3$. For $K=4$ (not shown) the qualitative behaviors are very similar to that of $K=3$. 
Again we see the instability of the BP equations for large $\beta$ and $q$ in case $K=2$. The discontinuity of the entropy for $K=2$ is again observed in the way that the energy density $e$ behaves with the overlap $q$. Compare it with the case $K=3$ where $e$ decreases monotonically by increasing $q$. Moreover, we see that when $K=2$ the overlap changes abruptly at $h=0$ for large $\beta$ which is a signal of clustering of the ground states in this case. In contrast, for $K=3$ the overlap smoothly approaches $1$ with increasing $\beta$ even at $h=0$ which is consistent with the absence of phase transitions in these systems.     

We end this section with an exact study of principal minors in matrices associated to fully connected graphs. In Appendix \ref{app:MF} we consider $N\times N$ matrices with elements $A_{ij}=\Lambda\delta_{i,j}-\Gamma(1-\delta_{i,j})$ in two scaling limits: (I) $\Lambda=N-1, \Gamma=1$, (II) $\Lambda=\mathrm{finite}, \Gamma=\gamma/N$ as $N\to \infty$. In the latter case we assume that $0\le \gamma \le \Lambda$ to represent positive and diagonally dominant matrices. We show that the behavior of the Laplacian of a complete graph, i.e., in the scaling limit (I), is very similar to that of Laplacian of random regular graphs of degree $K=4$. There are $N-1$ ground states with Hamming distances $2$ where only one of the indices is not present in the minor configuration. The scaling limit (II) however displays more interesting behaviors depending on the value of $\Lambda$. For $\Lambda>1$ there are $N-1$ ground states with only one index absent in the maximal minor configurations. On the other side, for $\Lambda<1$ there are $N-1$ ground states with only one index present the configurations. For $\Lambda=1$ all configurations have the same energy and the ground state entropy is maximal. Nevertheless, in both scaling limits we do not observe any finite-temperature phase transition; there is always a unique minimum of the free energy function which changes smoothly by increasing the inverse temperature $\beta$.

\begin{figure}
\includegraphics[width=16cm]{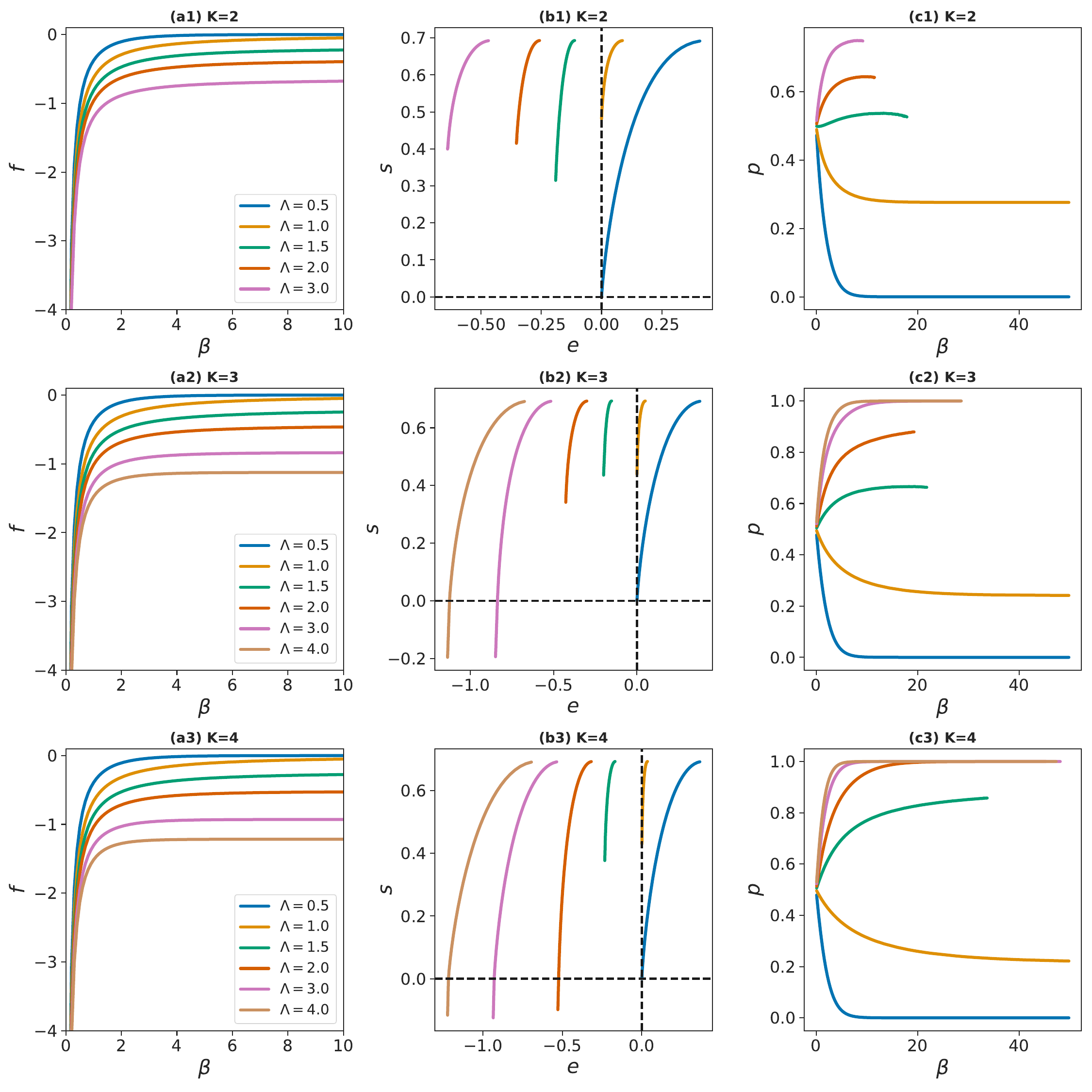} 
\caption{The asymptotic behavior of minors vs $\Lambda$ for quasi-Laplacian of random regular graphs of degree $K=2,3,4$. The free energy $f$, entropy vs energy $s(e)$, and probability of presence $p$ are reported for $\Lambda=0.5$ to $\Lambda=K+1$ in steps of size $\Delta\Lambda=0.5$. The quasi-Laplacian matrix is defined by diagonal elements $\Lambda$ and off-diagonal elements $-\Lambda/K$. The results are displayed for any $0<\beta<50$ as long as the entropy is concave and greater than $-0.2$. The parameters of the population dynamics algorithm are: population size $\mathcal{N}_p=10^4$, equilibration time $t_{eq}=10^6$, and averaging time $\Delta t_{avg}=10^4$.}\label{fig4}
\end{figure}

\begin{figure}
\includegraphics[width=16cm]{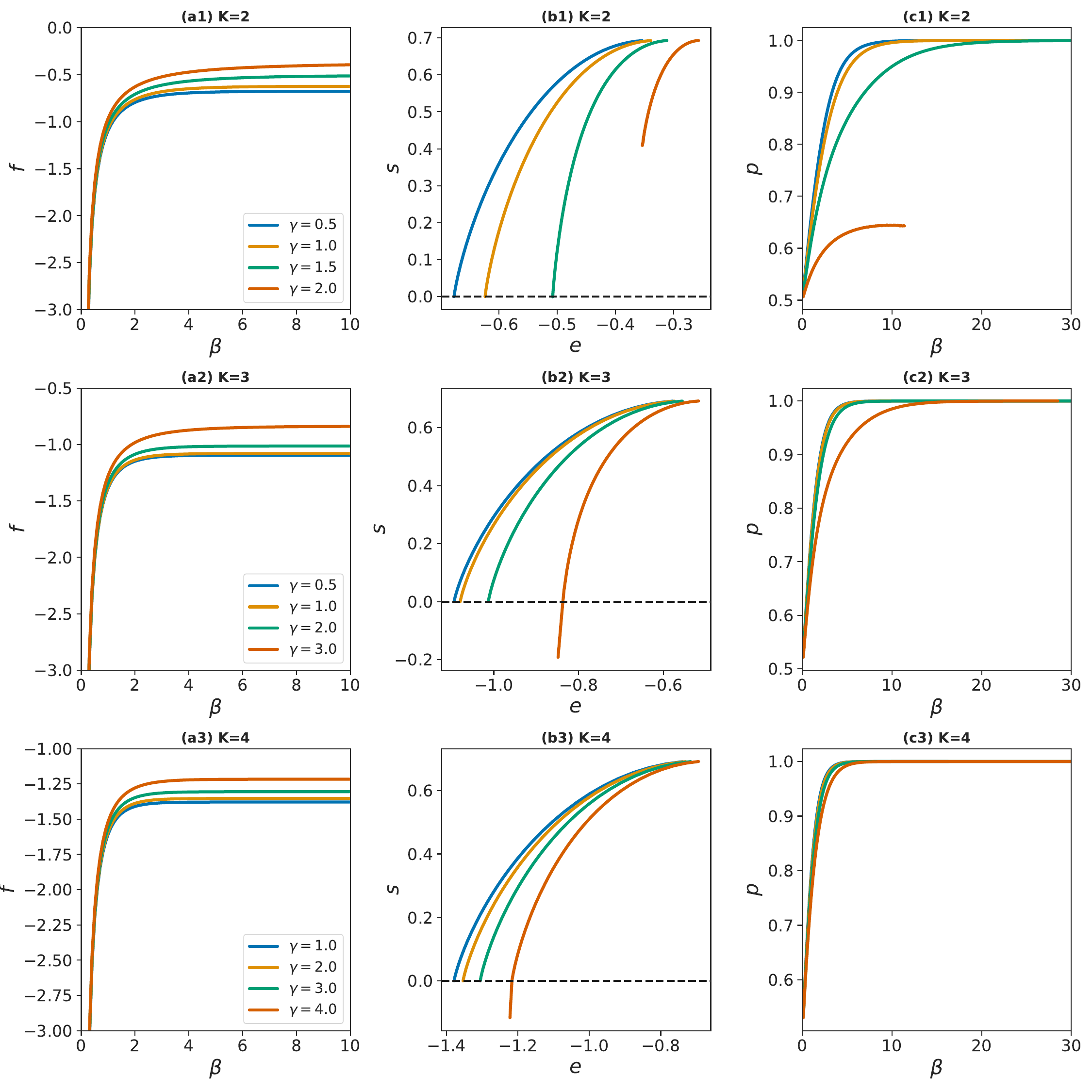} 
\caption{The asymptotic behavior of minors vs $\gamma$ for quasi-Laplacian of random regular graphs of degree $K=2,3,4$. The free energy $f$, entropy vs energy $s(e)$, and probability of presence $p$ are reported for $\gamma \in (0,K)$. The quasi-Laplacian matrix is defined by diagonal elements $K$ and off-diagonal elements $-\gamma/K$. The results are displayed for any $0<\beta<50$ as long as the entropy is concave and greater than $-0.2$. The parameters of the population dynamics algorithm are: population size $\mathcal{N}_p=10^4$, equilibration time $t_{eq}=10^6$, and averaging time $\Delta t_{avg}=10^4$.}\label{fig5}
\end{figure}

\begin{figure}
\includegraphics[width=10cm]{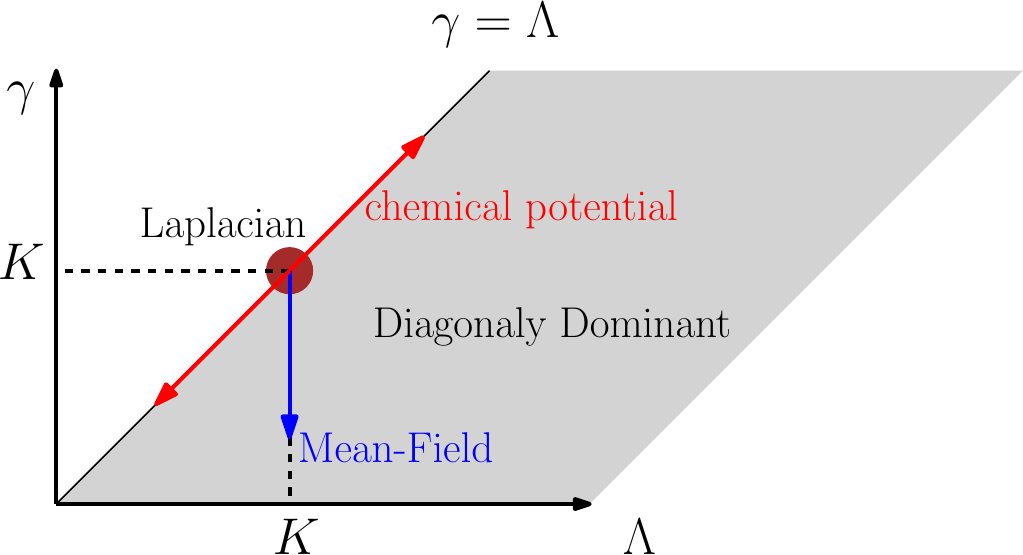} 
\caption{Deviation from the Laplacian in the region of diagonally dominant matrices. Moving in the diagonal direction $\gamma=\Lambda$ is like introducing a chemical potential for the size of minor configurations. The mean-field picture is recovered in the other direction $\gamma\to 0$.}\label{fig6}
\end{figure}

Figures \ref{fig4} and \ref{fig5} show how finite connectivity of random regular graphs changes the above mean-field picture.
We deviate from the Laplacian of random regular graphs in two different directions as depicted in Fig. \ref{fig6}.  
Let us start with the results displayed in Fig. \ref{fig4}. Here the diagonal matrix elements $\Lambda$ can be smaller or larger than the connectivity degree $K=2,3,4$ and the off-diagonal elements are $\Gamma=\Lambda/K$. This means that we are multiplying the Laplacian by $\Lambda/K$ which in turn modifies the value of minors by a factor $(\Lambda/K)^l$ for a minor configuration of size $l$. That is $-\ln(\Lambda/K)$ is like a chemical potential which now controls the number of present indices in the minors. We see that for $\Lambda\le 1$ the behavior of different degrees $K$ is very similar to that of the complete graphs except the smaller ground state entropy at $\Lambda=1$. Exact enumerations in small systems show that like the complete graphs, in this regime the limit $\beta\to \infty$ of the entropy density coincides with the zero-temperature entropy.  

For $\Lambda>1$, we observe numerical instabilities which can be attributed to the discontinuous behavior of the entropy density at zero temperature. Consider for instance the case $K=2$ (or a chain of length $N\gg 1$) where the energy of a connected cluster of $l$ present indices is given by $E(l)=-\ln(l+1)-l\ln(\Lambda/K)$. A sequence of clusters of present indices (represented by $1$s) which are separated by a single absent index (shown by $0$) make an ordered minor configuration for the chain graph; for instance, $\dots -0-111-0-111-0-111-0- \cdots$ with $l=3$ present indices in each cluster.
The energy density of such a minor configuration is given by $e(l)=-[\ln(l+1)+l\ln(\Lambda/K)]/(l+1)$.
This energy is minimized for a non-negative integer that is closer to $l^*=\frac{\Lambda}{K}e-1$. The optimal size of clusters increases linearly with $\Lambda$ starting from the all-zero minor configuration. A maximal minor configuration here can be considered as a close packing of clusters of effective size $l^*$ centered around the zeros (absent indices). As mentioned before, for the Laplacian ($K=2,\Lambda=2$) an optimal configuration is a dimer covering, or a close packing of (non-overlapping) rods of length $3$. Moreover, it is easy to see that the number of such optimal configurations is of order $N$ (by translation) so the entropy density at zero temperature is zero. For any finite cluster size $l$ the Hamming distances between the ground states is extensive but such states are not stable for any finite temperature; because the extensive entropy of excitations dominates the finite energies of the domain walls in this one-dimensional system. 

\begin{figure}
\includegraphics[width=14cm]{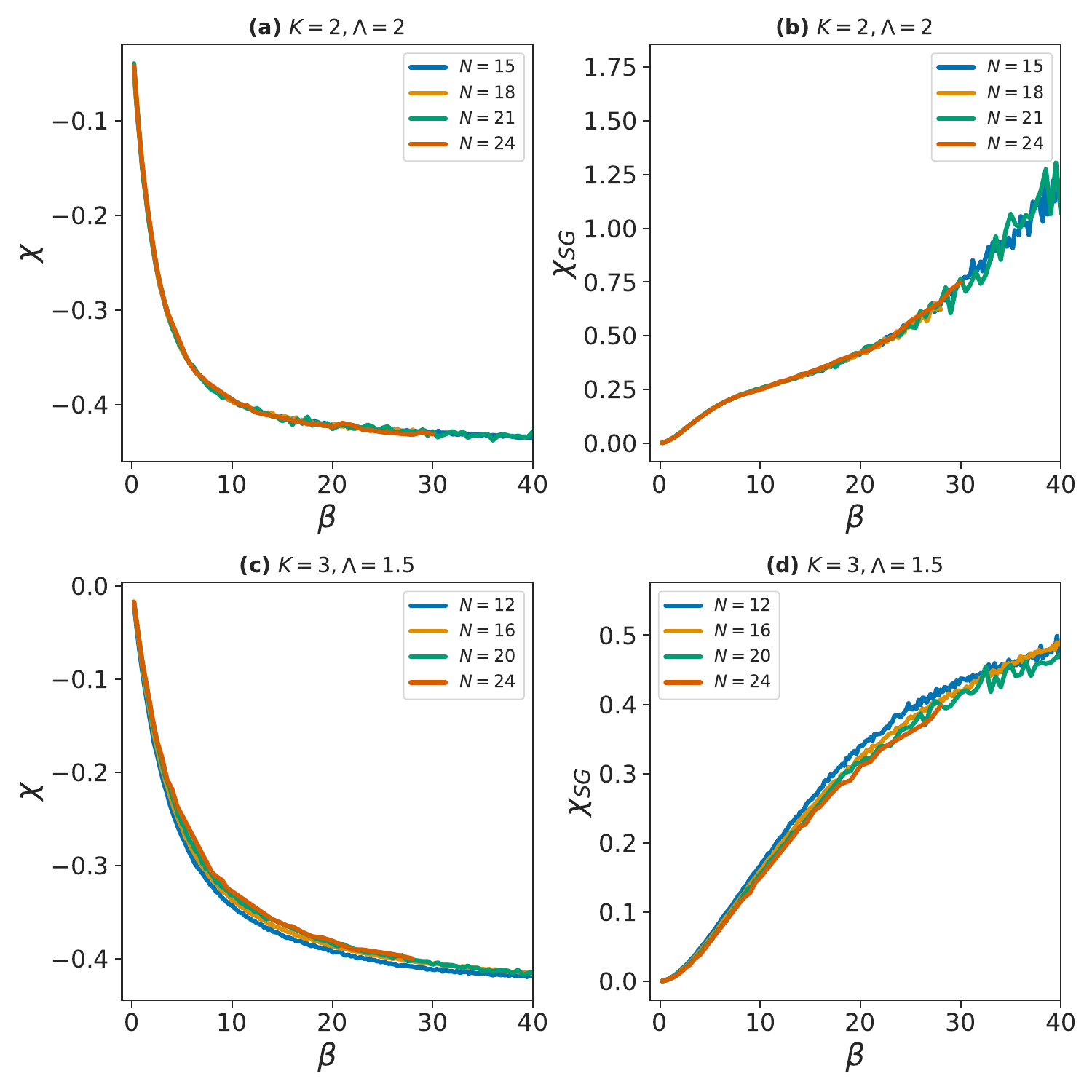} 
\caption{The sum of two-variable correlations when the entropy density displays a discontinuity. The ferromagnetic and spin-glass susceptibilities,  $\chi$ and $\chi_{SG}$ are exactly computed for random regular graphs of small sizes $N$. The matrices have diagonal elements $\Lambda$ and off-diagonal elements $-\Lambda/K$.  Panels ((a),(b)) show the results for random regular graphs of degree $K=2$ when $\Lambda=2$. Panels ((c),(d)) show the results for random regular graphs of degree $K=3$ when $\Lambda=1.5$.}\label{fig7}
\end{figure}

We see in Fig. \ref{fig4} that for degrees $K=3,4$ the main quantities change smoothly as long as $\Lambda\ge K$, where the number density of present indices in the optimal configurations is $1$. On the other hand, for $1<\Lambda <K$ we observe a discontinuous entropy density at zero temperature as $\Lambda$ approaches $1$, very similar to what happens for the smaller degree $K=2$. The ground states are random arrangements of an extensive number of zeros which are well separated on the interaction graph to maximize the number of spanning forests rooted at the zeros. The interval of $\Lambda$ values in which a numerical instability is displayed is of course reduced by increasing the degree $K$ approaching the mean-field limit. Figure \ref{fig7} displays the exact numerical results we obtain for the sum of two-variable correlations $\chi=\frac{1}{N}\sum_{i<j}[\langle \sigma_i\sigma_j\rangle-\langle \sigma_i\rangle\langle \sigma_j\rangle]$ and $\chi_{SG}=\frac{1}{N}\sum_{i<j}[\langle \sigma_i\sigma_j\rangle-\langle \sigma_i\rangle\langle \sigma_j\rangle]^2$ with $\sigma_i=2c_i-1=\pm 1$. The two susceptibilities $\chi$ and $\chi_{SG}$ are expected to diverge with $N$ near a magnetic or spin-glass phase transition, respectively \cite{marc-book-2009}. We observe that correlations remain short-ranged as $\beta$ increases in the two cases ($K=2,\Lambda=2$) and ($K=3,\Lambda=1.5$) which display a discontinuity in the entropy density.    

\begin{figure}
\includegraphics[width=16cm]{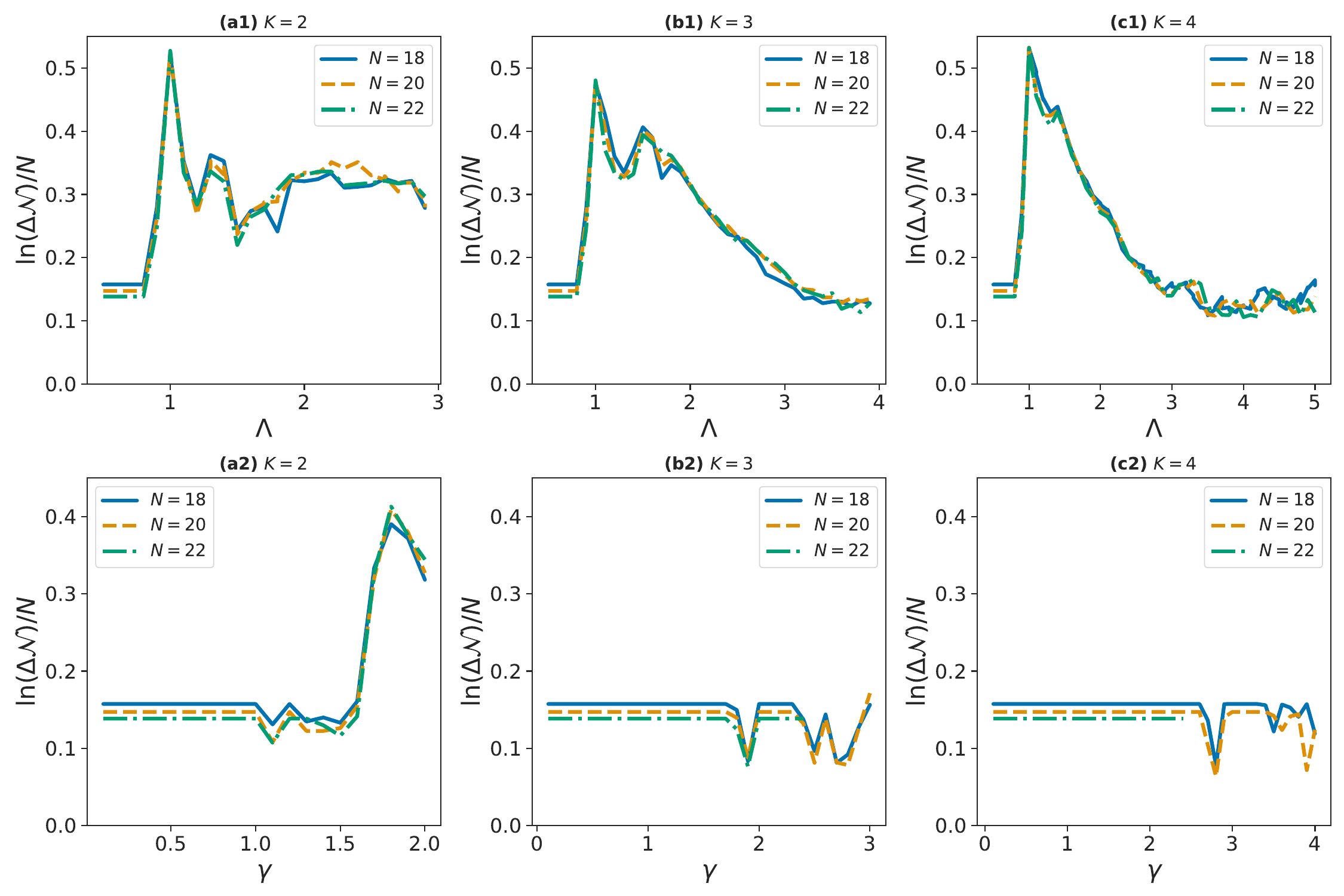} 
\caption{The gap in the number of minors $\Delta\mathcal{N}$ above the optimal minors in random regular graphs of degree $K=2,3,4$. The results are obtained by exhaustive enumeration for small sizes $N=18,20,22$. Panels ((a1),(b1),(c1)) show the changes with the chemical potential when the diagonal elements are $\Lambda$ and off-diagonal elements are  $-\Lambda/K$.
Panels ((a2),(b2),(c2)) show the variations in the mean-field direction when the diagonal elements are $K$ and off-diagonal elements are $-\gamma/K$. The curves start with horizontal lines which correspond to $\frac{1}{N}\ln(N)$.}\label{fig8}
\end{figure}

Figure \ref{fig5} shows the results for matrices with diagonal elements $K$ and off-diagonal elements $-\gamma/K$ when $\gamma \in (0,K)$. By decreasing the magnitude of the off-diagonals compared to the diagonals we approach the mean-field behavior. Here the picture is simpler, with curves that smoothly approach (up to very large $\beta$) the presence probability $p=1$, where nearly all the indices are present in the optimal minor configurations. In Fig. \ref{fig8} we report exact numerical results for variations in the number of minor configurations just above the optimal ones in small systems of sizes $N=18,20,22$. We see that the entropy gap remains nonzero (for $K=2$) or approaches zero (for $K=3,4$) when the number of present indices changes by deviating from the Laplacian in the diagonal direction of Fig. \ref{fig6}. Again we observe that the entropy density displays a discontinuity for $K=2$, and for $K=3,4$ when $\Lambda$ is close to $1$. On the other hand, the entropy gap tends to zero when we approach the mean-field limit, except for $K=2$ and $\gamma$ close to $K$.

\subsection{Zero-temperature limit}\label{S23}
At zero temperature the Gibbs probability measure is concentrated on the ground states of the system with minimum energy $E=-\ln\mathrm{det}\mathbf{A}(\mathbf{c})$, i.e., the maximal minors of the matrix $\mathbf{A}$.
To study the ground states we take the limit $\beta \to \infty$ of the higher-level BP equations.
Let us assume that $M_{i\to j}(c_i,v_{i\to j})=e^{\beta w_{i\to j}(c_i,v_{i\to j})}$ as $\beta \to \infty$. Then, from Eq. \ref{BPM} we obtain the so called MaxSum (MS) equations \cite{baldassi-polito-2009} for the cavity messages $w_{i\to j}(c_i,v_{i\to j})$, see also Appendix \ref{app:bp},
\begin{align}
w_{i\to j}(c_i,v_{i\to j})= -2\Delta g_{i\to j}
+\max_{\{c_k,v_{k\to i}\:k\in \partial i\setminus j\}: \mathbb{I}(v_{i\to j})} \sum_{k\ne i,j} w_{k\to i}(c_k,v_{k\to i})-C_{i\to j}.
\end{align}
The constant $C_{i\to j}$ is chosen such that $\max_{c_i,v_{i\to j}} w_{i\to j}(c_i,v_{i\to j})=0$.

To solve the above equations by iteration we use a discrete representation of the variances $v_{i\to j}=n_{i\to j}\delta v$ with integers $n_{i\to j}$ for a small $\delta v$. The maximum over the variables $\{c_k,v_{k\to i}\:k\in \partial i\setminus j\}$ can efficiently be computed by using repeated convolutions of the messages $w_{k\to i}$. In this way, the time complexity of each iteration of the algorithm in a graph of degree $K$ is of order $N(KN_v)^2$, where $N_v$ is the number of the possible values of discrete variances. In each iteration all the cavity messages are updated in a random and sequential way. The number of iterations needed to solve the equations is of order $100$, independent of the problem size $N$.

In the same way, we can take the limit $\beta \to \infty$ of the free energies in Eqs. \ref{BPF},
\begin{align}
\Delta f_i=-\max_{c_i}\left( \max_{\{c_j,v_{j\to i}:j\in \partial i\}} \left(\sum_{j\in \partial i} w_{j\to i}(c_j,v_{j\to i})-2\Delta g_{i} \right) \right),
\end{align}
and
\begin{align}
\Delta f_{ij}=-\max_{\{s_{ij},v_{ij}\}} \left( w_{i\to j}(c_i,v_{i\to j})+w_{j\to i}(c_j,v_{j\to i})-2\Delta g_{ij} \right).
\end{align}
These equations are useful to obtain an estimation of the ground state energy within the Bethe approximation.

In practice, we use a reinforcement algorithm to find a minimum energy configuration of the system \cite{alfredo-prl-2006}. The idea is to reinforce the MS equations by introducing a bias feedback to the equations. The bias is provided by the local MS messages to gradually converge the algorithm towards a ground state of the system. More precisely, the reinforced MS equations at iteration $t+1$ read as follows:
\begin{align}
w_{i\to j}^{t+1}(c_i,v_{i\to j})=r(t)w_i^t(c_i)-2\Delta g_{i\to j}
+\max_{\{c_k,v_{k\to i}\:k\in \partial i\setminus j\}: \mathbb{I}(v_{i\to j})} \sum_{k\ne i,j} w_{k\to i}^t(c_k,v_{k\to i})-C_{i\to j}.
\end{align}
with the initial MS messages $w_{i\to j}^0=w_i^0=0$. The reinforcement parameter $r(t)$ increases linearly with time as $r(t+1)=r(t)+\delta r$, starting from $r(0)=0$ with $\delta r\ll 1$. Similarly, the local MS messages are given by  
\begin{align}
w_i^{t+1}(c_i)=r(t)w_i^t(c_i)
+\max_{\{c_j,v_{j\to i}:j\in \partial i\}} \left(\sum_{j\in \partial i} w_{j\to i}^t(c_j,v_{j\to i})-2\Delta g_{i} \right).
\end{align}
These messages then provide a candidate for the minimum energy configuration
\begin{align}
c_i^*=\arg\max w_i^{t}(c_i).
\end{align}
Note that the MaxSum algorithm is not expected to work well in a loopy graph but the algorithm can always be used to provide a candidate for the optimal configuration and therefore an upper bound for the minimum energy even in a loopy interaction graph.

\begin{figure}
\includegraphics[width=14cm]{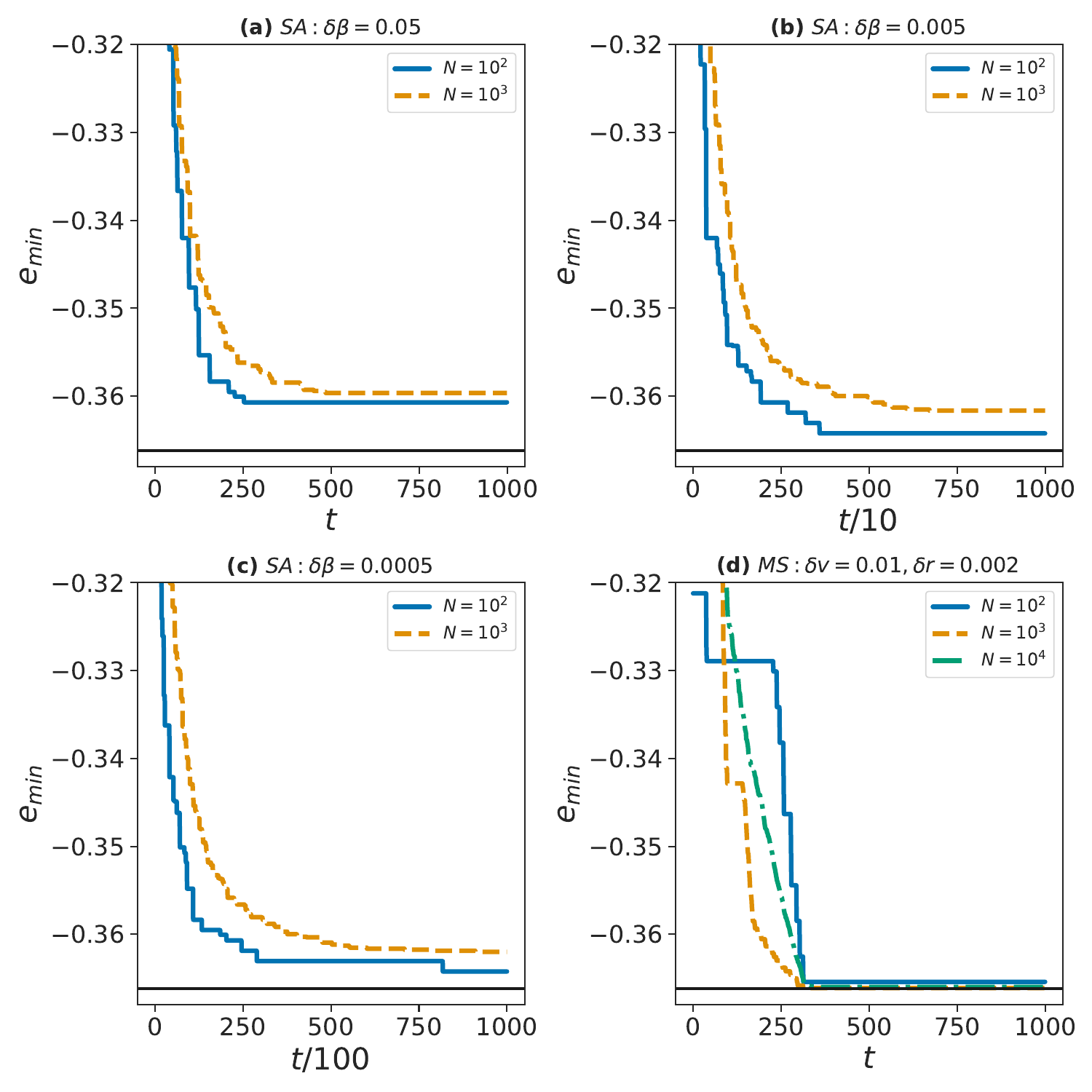} 
\caption{Finding a maximal minor of Laplacian of a closed chain of size $N$ by an optimization algorithm. The results are obtained by the simulated annealing (SA) and MaxSum (MS) algorithms with the indicated parameters. The horizontal lines show the exact value expected in the thermodynamic limit. The ground states are dimer coverings of energy density $e_0=-\ln(3)/3$. Here $t$ is the number of iterations in the algorithms. In each iteration all the variables are updated once.}\label{fig9}
\end{figure}

\begin{figure}
\includegraphics[width=16cm]{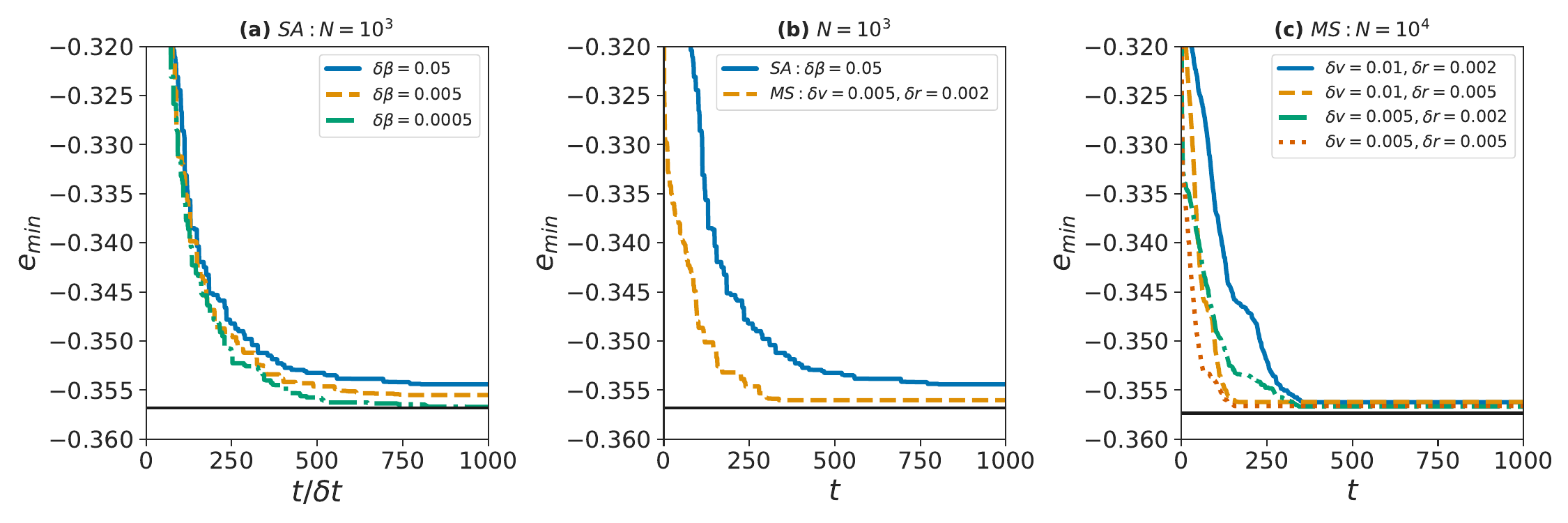} 
\caption{Finding a maximal minor of a random Laplacian of a closed chain of size $N$ by an optimization algorithm. The results are obtained by the simulated annealing (SA) and MaxSum (MS) algorithms with the indicated parameters. The off-diagonal elements of the matrices are uniformly distributed in $(-2,0)$ and the diagonal elements in each row are minus the sum of other elements in that row. The horizontal lines show the numerical value predicted by the MaxSum equations. The ground states are random configurations close to the dimer coverings of the ordered system. The number of iterations in the SA algorithm is scaled with $\delta t=1,10,100$ for $\delta\beta=0.05,0.005,0.0005$, respectively.}\label{fig10}
\end{figure}

We start by applying the above algorithm to Laplacian of a closed chain where we know the ground state energy and configurations. Recall that the maximal minors of a chain are dimer coverings which are separated by extensive Hamming distances in the configuration space. We will see that this nontrivial structure of the ground states can make it difficult for a local optimization algorithm to find an optimal solution of the problem. In fact, an algorithm may try to minimize energy of the system by constructing a configuration that is locally like one of the ground states of the system. If these states are very different then it could be very difficult to overcome these energy/entropy barriers and to end up with one of the optimal states of the system. Figure \ref{fig9} shows how the minimum energy density suggested by the reinforced MS algorithm approaches the theoretical value $e_0=-\ln(3)/3$ expected for a very large chain. For comparison, we also report the results obtained by a simulated annealing (SA) algorithm; in each step, one variable $c_i$ is selected and its value is changed with probability $\min\{1,e^{-\beta\Delta E}\}$, where $\Delta E$ is the resulted change in the energy function. In one iteration of the algorithm all variables are selected in a random and sequential way, as in the MS algorithm. We start from a high temperature (small $\beta$) and in each iteration increase the inverse temperature by $\delta\beta$ to approach a low-energy configuration. We see in Fig. \ref{fig9} that it is more difficult for the SA algorithm to find a ground state than the MS algorithm. A similar behavior is observed in Fig. \ref{fig10} for random Laplacian of a closed chain. Here the off-diagonal elements are random and independent numbers uniformly distributed in $(-2,0)$ to have the average $-1$ as in the ordered case. The diagonal elements in each row are minus the sum of the off-diagonal elements in that row, like a Laplacian. Note that the MS algorithm finds a good estimation of the minimum energy in a smaller number of iterations compared to the SA algorithm. Moreover, a single iteration of SA algorithm is more time consuming than the MS algorithm because we need to compute the matrix determinants for each update of the configuration. This does not allow us in practice to study larger systems by the simulated annealing algorithm.

\begin{figure}
\includegraphics[width=14cm]{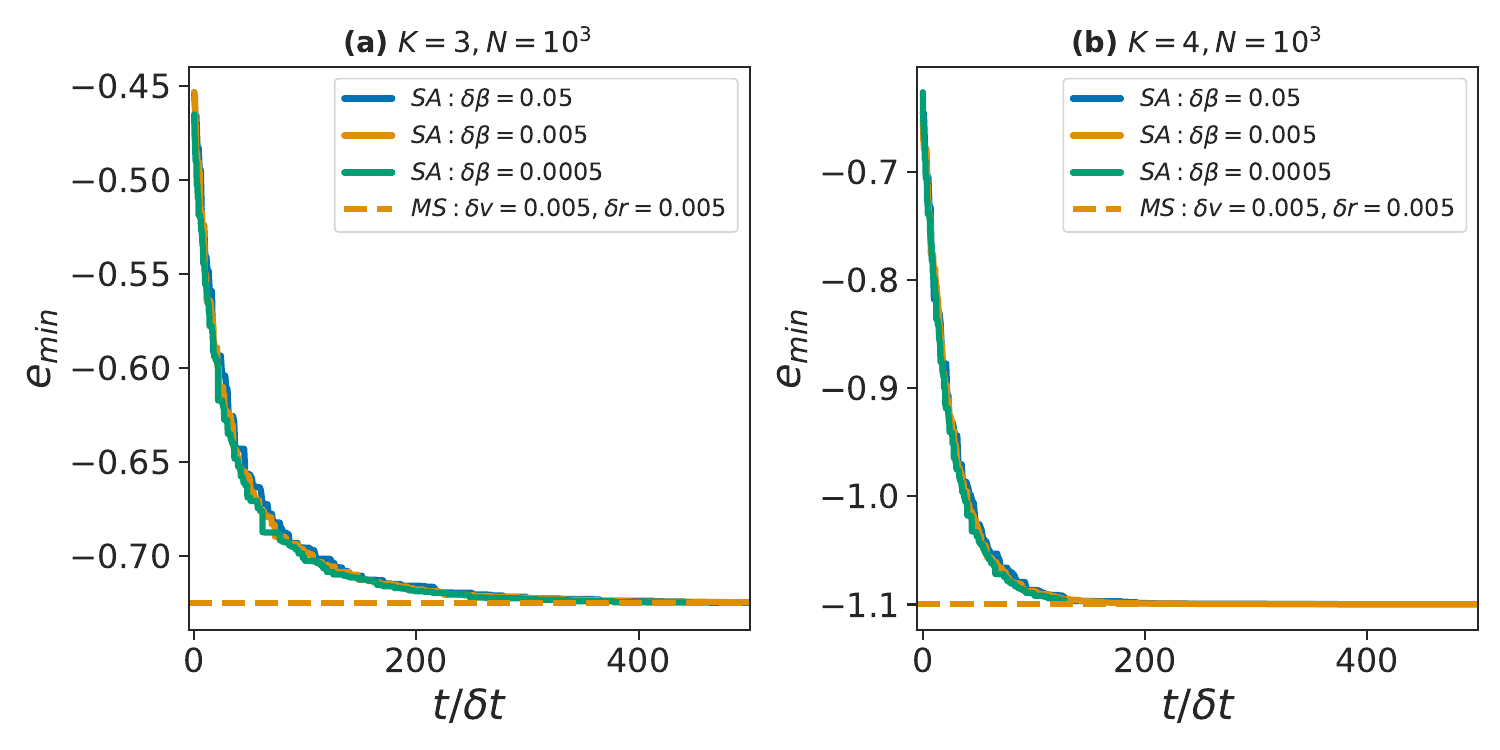} 
\caption{Finding a maximal minor of a random Laplacian of random regular graphs of degree $K=3,4$ by an optimization algorithm. The results are obtained by the simulated annealing (SA) and MaxSum (MS) algorithms with the indicated parameters. The off-diagonal elements of the matrices are uniformly distributed in $(-2,0)$ and the diagonal elements in each row are minus the sum of other elements in that row. The horizontal lines show the minimum energies obtained by the MaxSum algorithm in a few hundred iterations. The fraction of present indices are around $p=0.88 (K=3)$ and $p=0.99 (K=4)$. The number of iterations in the SA algorithm is scaled with $\delta t=1,10,100$ for $\delta\beta=0.05,0.005,0.0005$, respectively.}\label{fig11}
\end{figure}

Figure \ref{fig11} displays the results we obtain for random Laplacian of random regular graphs of degree $K=3,4$. The random Laplacians are constructed as described in the previous paragraph. We observe that the SA algorithm is more effective for large connectivity degrees compared to the case of chain ($K=2$) because the energy landscape is simpler for $K>2$. For the same reason, the MS algorithm very quickly converges to the ground state of these systems. The optimal states of random Laplacians are indeed very close to the ground states of the ordered systems with only a small fraction of indices not present in the optimal states. The ground states in the ordered versions of these systems are minor configurations with nearly all indices, except one or two of them, present in the configuration. For instance, only one of the indices is not present in the ground states of the Laplacian of random regular graphs of degree $K=4$. There are therefore $N$ ground states in such a homogeneous system and the Hamming distance between two optimal configurations is $2$. This trivial structure of the ground states makes it also easy for a local optimization algorithm like SA to find a ground state.

\section{Conclusion and Discussion}\label{S3}
In summary, a Gaussian representation of minors for symmetric, positive, and diagonaly-dominant (DD) matrices was employed to estimate the free energy, entropy, and energy of relevant minors at different temperatures by the Bethe approximation. The estimation is expected to be asymptotically exact for matrices which have a locally tree-like graph representation. Specifically, we studied the energy and entropy landscape of the Laplacian of random regular graphs of degree $K=2,3,4$. For large degrees $K=3,4$ the interesting quantities change smoothly by increasing the inverse temperature $\beta$. The case $K=2$, however, displays a discontinues entropy and numerical instability close to this discontinuity which separates the zero- and finite-temperature behaviors. Nevertheless, we do not observe a finite-temperature phase transition in these systems for $K=2,3,4$. The same result is obtained by an exact treatment of principal minors in DD matrices which are defined by homogeneous fully-connected graphs.    

The zero-temperature limit of the Bethe equations is used as an optimization (MaxSum) algorithm to find an estimation of the ground state energy and optimal configurations (maximal minors) of random DD matrices. Here the MaxSum algorithm is more efficient than the standard simulated annealing algorithm regarding the number of iterations that are needed to find a good estimation of the ground state(s). The time complexity of the MaxSum algorithm is proportional to $N$ in a finite-connectivity graph which is represented by a sparse $N\times N$ matrix. 

The first part of the study enables us to estimate the sum of powers of principal minors which is relevant to computation of the partition function and the Shannon-R\'enyi entropy of quantum systems such as the Hubbard model and the
transverse field Ising model \cite{morteza-2023}. In the second part, we obtain an approximate message-passing optimization algorithm which can be applied to sampling problems where a subset of maximally independent configurations are needed as in the determinantal point processes.  

It seems that restriction to diagonally dominant matrices is the main reason behind the absence of finite-temperature phase transitions in such systems. This is the case also for positive but quasi one-dimensional matrices where entropy is dominated for any finite temperature. It would be interesting to investigate the nature of the possible phase transitions for an arbitrary positive matrix. If the problem of finding the optimal minors is in general a computationally hard problem, then one expects to observe spin-glass phases in these systems. In this study, we were mainly focused on regular quasi-Laplacian matrices with no randomness in matrix elements. A more detailed investigation of random and diagonally dominant matrices is needed to characterize the phase diagram of this class of matrices. 

Recall that the principal minors of a Laplacian give the number of possible spanning forests that are generated by a set of trees rooted at the zeros of the minor configuration (the absent indices). By maximizing the number of such forests we are indeed increasing the strength of an effective repulsive interaction between the zeros of the minor configuration. In this study we used a chemical potential to control the expected number of roots (zeros) in the relevant minors. We showed that a maximal minor in a chain ($K=2$) is a close packing of the roots with an effective distance $l^*$ which is determined by the chemical potential. It would be interesting to investigate the statistical properties of this system of repulsive zeros, for example in a Hamming space from the perspective of the coding theory.

\acknowledgments
MAR thanks CNPq and FAPERJ (grant number E-26/210.062/2023) for partial support.
This work was performed using the Academic Leiden Interdisciplinary Cluster Environment (ALICE), the compute resources provided by Leiden University.

\appendix
\numberwithin{equation}{section}
\counterwithin{figure}{section}

\section{Bethe approximation: Belief Propagation algorithm}\label{app:bp}
In this section we briefly describe the Bethe approximation and the resulting Belief Propagation (BP) algorithm for an Ising model defined on an interaction graph $G$. Consider $N$ Ising variables of configurations $\mathbf{s}=\{s_i=\pm 1:1,\dots,N\}$ and energy function $E[\mathbf{s}]=-\sum_{(ij)}J_{ij}s_is_j-\sum_i h_i s_i$. Note that the interaction term is a sum over edges $(ij)$ of the interaction graph $G$. In thermal equilibrium at inverse temperature $\beta$, the Boltzmann factor $e^{-\beta E[\mathbf{s}]}/Z$ gives the statistical weight of configurations $\mathbf{s}$, where the partition function $Z=\sum_{\mathbf{s}}e^{-\beta E[\mathbf{s}]}$ is a normalization constant. Here we are interested in the marginal probabilities of local variables $\mu_i(s_i)$ at equilibrium. In limit $\beta \to \infty$ these marginals can be used to find the ground state(s) of the system.

\begin{figure}
\includegraphics[width=8cm]{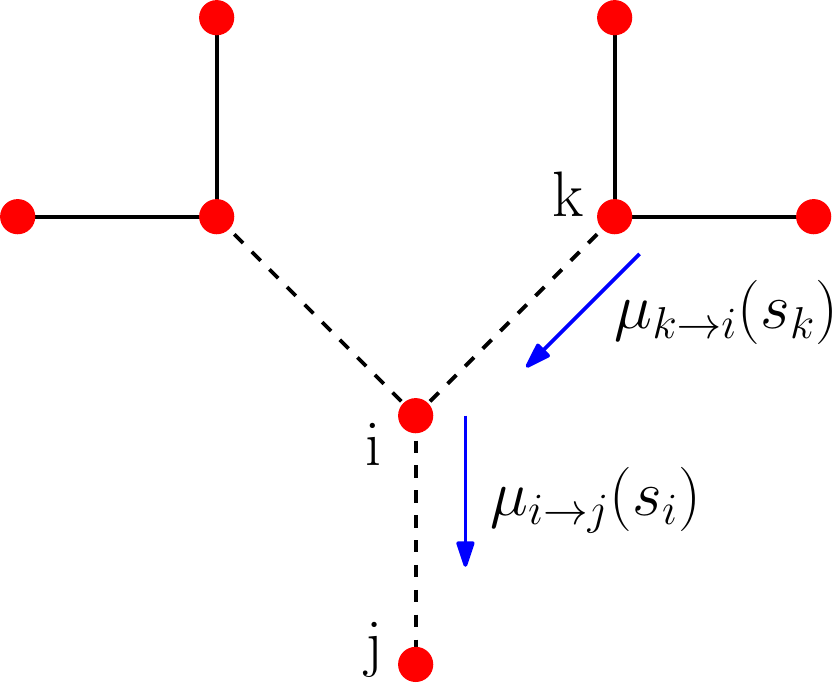} 
\caption{An illustration of the Belief Propagation algorithm in a tree interaction graph. The node variable $i$ sends the cavity message $\mu_{i\to j}(s_i)$ to node $j$ depending on the messages $\mu_{k\to i}(s_k)$ that it receives from the other neighbors.}\label{fig12}
\end{figure}

Bethe equations are indeed recursive equations for the cavity marginals $\mu_{i\to j}(s_i)$, see Fig. \ref{fig12}. This is the probability of state $s_i$ for variable $i$ in the absence of interaction with variable $j$, i.e., when the interaction term $-J_{ij}s_is_j$ is removed from the energy function. It is also assumed that in the absence of this interaction $s_i$ is independent of the state of the other neighbors of variable $j$. This assumption is valid when the interaction graph $G$ is a tree, or has very large loops such that locally it is like a tree. Then, the Bethe equations for the cavity marginals read
\begin{align}
\mu_{i\to j}(s_i)\propto e^{\beta h_i s_i}\prod_{k\in \partial i\setminus j}\left(\sum_{s_k}e^{\beta J_{ik}s_is_k}\mu_{k\to i}(s_k)\right),
\end{align}
were $\partial i$ is the set of neighbors of variable $i$ in the interaction graph. The normalization constant is obtained from $\sum_{s_i}\mu_{i\to j}(s_i)=1$. These equations can be solved by iteration starting from random initial messages $\mu_{i\to j}(s_i)$ and updating the messages according to the above equations. Having the cavity marginals one finds the local marginals $\mu_i(s_i)$ by considering the effects of all neighboring variables,
\begin{align}
\mu_{i}(s_i)\propto e^{\beta h_i s_i}\prod_{k\in \partial i}\left(\sum_{s_k}e^{\beta J_{ik}s_is_k}\mu_{k\to i}(s_k)\right).
\end{align}
Moreover, an estimation of the free energy $F=-\frac{1}{\beta}\ln Z$ can be obtained from the cavity marginals. To this end, we need to compute the contributions of all variables and interactions in the partition function. Assuming that the interaction graph is a tree, the free energy is given by
\begin{align}
F=\sum_i \Delta F_i-\sum_{(ij)}\Delta F_{ij},
\end{align}
This is obtained by writing a recursive equation for the partition function starting from an arbitrary variable \cite{baldassi-polito-2009}. The local free energy changes are related to the cavity marginals as follows:
\begin{align}
e^{-\beta \Delta F_i} &=\sum_{s_i}e^{\beta h_i s_i}\prod_{k\in \partial i}\left(\sum_{s_k}e^{\beta J_{ik}s_is_k}\mu_{k\to i}(s_k)\right),\\
e^{-\beta \Delta F_{ij}} &=\sum_{s_i,s_j}e^{\beta J_{ij}s_is_j}\mu_{i\to j}(s_i)\mu_{j\to i}(s_j).
\end{align}

To obtain some information about the ground states, we take the limit $\beta\to \infty$ and assume that $\mu_{i\to j}(s_i)=e^{\beta m_{i\to j}(s_i)}$. The BP equations in this limit (called MaxSum equations) are 
\begin{align}
m_{i\to j}(s_i)=h_i s_i+\sum_{k\in \partial i\setminus j}\max_{s_k}(J_{ik}s_is_k+m_{k\to i}(s_k))-C_{i\to j},
\end{align}
where $C_{i\to j}$ is obtained by normalization condition $\max_{s_i} m_{i\to j}(s_i)=0$. As before, the equations are solved by iteration. It is useful for a better convergence to introduce reinforcement in the algorithm; at each step of the iteration algorithm a bias field (feedback) is added to the system to favor the more probable states of the variables. More precisely, the reinforced MaxSum equations at time step $t+1$ read as follows,
\begin{align}
m_{i\to j}^{t+1}(s_i)=h_i s_i+r(t)m_i^t(s_i)+\sum_{k\in \partial i\setminus j}\max_{s_k}(J_{ik}s_is_k+m_{k\to i}^t(s_k))-C_{i\to j}.
\end{align}
Similarly, one obtains the local messages,
\begin{align}
m_{i}^{t+1}(s_i)=h_i s_i+r(t)m_i^t(s_i)+\sum_{k\in \partial i}\max_{s_k}(J_{ik}s_is_k+m_{k\to i}^t(s_k))-C_{i}.
\end{align}
The reinforcement parameter $r(t)$ is zero at the beginning of the algorithm and increases slowly by the number of iterations.
At each time step, one can find an approximate ground state of the system by computing the maximal states of the local messages, that is
\begin{align}
s_i^*=\arg\max_{s_i} m_{i}^{t}(s_i).
\end{align}
For more details and applications of the BP and MaxSum algorithms in other problems see \cite{baldassi-polito-2009}.

\section{The population dynamics algorithm}\label{app:pop}

In this section we describe the population dynamics that is used to solve the higher-level BP equations
in the Gaussian representation of the minors,
\begin{align}
M_{i\to j}(c_i,v_{i\to j}) \propto \prod_{k\in \partial i\setminus j}\left( \sum_{c_k}\int dv_{k\to i} M_{k\to i}(c_k,v_{k\to i}) \right)
\times \mathbb{I}(v_{i\to j})e^{-2\beta \Delta g_{i\to j}}.
\end{align}
Recall that the variances $v_{i\to j}$ are solutions to the BP equations for the Gaussian variables,
\begin{align}\label{app-vij}
\frac{1}{v_{i\to j}}=(1-c_i)+c_iA_{ii}-\sum_{k\ne i,j} c_iA_{ik}^2c_k v_{k\to i},
\end{align}
and
\begin{align}\label{app-gij}
2\Delta g_{i\to j}=\ln(2\pi v_{i\to j}).
\end{align}
 
The probability distributions $M_{i\to j}(c_i,v_{i\to j})$ are represented by populations of the variables $P_{i\to j}=\{(c_i^a,v_{i\to j}^a):a=1,\dots,\mathcal{N}_p\}$ on each directed link $(i\to j)$. In addition, we introduce populations of the Boltzmann weights $W_{i\to j}=\{e^{-2\beta \Delta g_{i\to j}^a}:a=1,\dots,\mathcal{N}_p\}$ which are associated to members $(c_i^a,v_{i\to j}^a)$ of the $P_{i\to j}$. More about the population dynamics in general can be found in \cite{lenka-thesis}.

At the beginning of the algorithm we set the initial values $v_{i\to j}^a=0$, $c_i^a=0,1$ (with equal probability) and $\Delta g_{i\to j}^a=\infty$. In each iteration of the population dynamics we go through all the directed links $(i\to j)$ in a random and sequential way and do the following steps:

\begin{itemize}

\item select a member $(c_k^{a_k},v_{k\to i}^{a_k})$ from $P_{k\to i}$ for $k\in \partial i\setminus j$; 

\item compute $v_{i\to j}(c_i)$ and $\Delta g_{i\to j}(c_i)$ (Eqs. \ref{app-vij} and \ref{app-gij}) for $c_i=0,1$ given the messages $(c_k^{a_k},v_{k\to i}^{a_k})$; 

\item select a member $(c_i^{a_0},v_{i\to j}^{a_0})$ of $P_{i\to j}$ and replace it with $(0,v_{i\to j}(0))$ with probability $e^{-2\beta(\Delta g_{i\to j}^{a_0}-\Delta g_{i\to j}(0))}$. If accepted replace $\Delta g_{i\to j}^{a_0}$ with $\Delta g_{i\to j}(0)$;

\item select a member $(c_i^{a_1},v_{i\to j}^{a_1})$ of $P_{i\to j}$ and replace it with $(1,v_{i\to j}(1))$ with probability $e^{-2\beta(\Delta g_{i\to j}^{a_1}-\Delta g_{i\to j}(1))}$. If accepted replace $\Delta g_{i\to j}^{a_1}$ with $\Delta g_{i\to j}(1)$; 

\end{itemize}

Note that the members of the populations are selected randomly and uniformly. The updates are repeated for $t_{eq}$ iterations to reach a steady state where the average quantities are stationary.

In the stationary state of the population dynamics, we obtain an estimation of the free energy
\begin{align}
Nf_g=\sum_i \Delta f_i-\sum_{i<j} \Delta f_{ij},
\end{align}
with local free energies that are given by
\begin{align}
e^{-\beta \Delta f_i} &=\sum_{c_i} \prod_{j\in \partial i}\left( \sum_{c_j}\int dv_{j\to i} M_{j\to i}(c_j,v_{j\to i}) \right) e^{-2\beta \Delta g_i},\\
e^{-\beta \Delta f_{ij}} &=\sum_{c_i,c_j} \int dv_{ij} e^{-2\beta \Delta g_{ij}} M_{i\to j}(c_i,v_{i\to j}) M_{j\to i}(c_j,v_{j\to i}). 
\end{align}
Here $v_{ij}=\{v_{i\to j},v_{j\to i}\}$. Recall that
\begin{align}\label{app-gi}
2\Delta g_i &=\ln(2\pi v_i),\\
2\Delta g_{ij} &=-\ln\left(v_{i\to j}v_{j\to i}\mathrm{det}(B(ij))\right),
\end{align}
where
\begin{align}
\frac{1}{v_i} &=(1-c_i)+c_iA_{ii}-\sum_{k\ne i} c_iA_{ik}^2c_k v_{k\to i},\\
B(ij) &=\begin{pmatrix} \frac{1}{v_{i\to j}} & c_iA_{ij}c_j \\ c_iA_{ij}c_j & \frac{1}{v_{j\to i}} \end{pmatrix}.
\end{align}

To compute the $\Delta f_i$ we repeat the following steps for $\Delta t_{avg}$ times: 

\begin{itemize}

\item select a member $(c_k^{a_k},v_{k\to i}^{a_k})$ from $P_{k\to i}$ for $k\in \partial i$; 

\item compute $\Delta g_{i}(c_i)$ (Eq. \ref{app-gi}) and $w_i(c_i)\equiv e^{-2\beta\Delta g_{i}(c_i)} $ for $c_i=0,1$ given the messages $(c_k^{a_k},v_{k\to i}^{a_k})$; 

\end{itemize}

From the above computation we obtain the averages $Z_i(c_i)\equiv \langle w_i(c_i)\rangle$, and $\Delta E_i(c_i)\equiv 2\langle w_i(c_i)\Delta g_{i}(c_i) \rangle$. Then $\Delta f_i=-\ln(Z_i(0)+Z_i(1))/\beta$, the probability of $c_i=1$ is $p_i=Z_i(1)/(Z_i(0)+Z_i(1))$, and we define $\Delta e_i=(\Delta E_i(0)+\Delta E_i(1))/(Z_i(0)+Z_i(1))$. 

In the same way, to compute the $\Delta f_{ij}$ we repeat the following steps for $\Delta t_{avg}$ times: 

\begin{itemize}

\item select $(c_i^{a},v_{i\to j}^{a}),(c_j^{b},v_{j\to i}^{b})$ from $P_{i\to j}$ and $P_{j\to i}$; 

\item compute $\Delta g_{ij}$ (Eq. \ref{app-gi}) and $w_{ij}\equiv e^{-2\beta\Delta g_{ij}} $ given the above messages; 

\end{itemize}

From the above computation we obtain the averages $Z_{ij}\equiv \langle w_{ij}\rangle$, and $\Delta E_{ij}\equiv 2\langle w_{ij}\Delta g_{ij} \rangle$. Then $\Delta f_{ij}=-\ln(Z_{ij})/\beta$ and we define $\Delta e_{ij}=\Delta E_{ij}/Z_{ij}$.

Finally, the free energy density $f$ and energy density $e$ are obtained
\begin{align}
Nf=\sum_i \Delta f_i-\sum_{i<j} \Delta f_{ij}-N\ln(2\pi),\\
Ne=\sum_i \Delta e_i-\sum_{i<j} \Delta e_{ij}-N\ln(2\pi).
\end{align}

\section{The Bethe equations for random regular graphs}\label{app:rrg}

Consider the Laplacian of a random regular graph of degree $K$. Here by symmetry all nodes are equivalent. Thus the BP equations
for the cavity marginals $M_{i\to j}(c_i,v_{i\to j})$ of the Gaussian representation 
\begin{align}
M_{i\to j}(c_i,v_{i\to j}) \propto \prod_{k\in \partial i\setminus j}\left( \sum_{c_k}\int dv_{k\to i} M_{k\to i}(c_k,v_{k\to i}) \right)
\times \mathbb{I}(v_{i\to j})e^{-2\beta \Delta g_{i\to j}},
\end{align}
reduce to a single equation for $M_{\to}(c,v)$ which does not depend on the indices $i$ and $i\to j$.
We take $M_{\to}(c,v)=(1-p)\delta(v-1)\delta_{c,0}+p\rho(v)\delta_{c,1}$ with $p$ for the probability of presence of index $i$ in the minor configuration. We recall that for $c=0$ the variance is $v=1$ independent of the states of the neighbors. The probability distribution of $v$ for $c=1$ is shown by $\rho(v)$.
Then the BP equations for $p$ and $\rho(v)$ read
\begin{align}
p &=\frac{z_1(K-1)}{z_0(K-1)+z_1(K-1)},\\
\rho(v) &=\frac{z(v:K-1)}{z_1(K-1)},
\end{align}
where $z_0(K-1) =e^{-\beta\ln(2\pi)}$,
\begin{align}
z_1(K-1) =\sum_{l=0}^{K-1}C(l,K-1)p^{l}(1-p)^{K-1-l}
\int \prod_{k=1}^ldv_k\rho(v_k)e^{-\beta\ln(\frac{2\pi}{K-\sum_kv_k})},
\end{align}
and
\begin{align}
z(v:K-1) =\sum_{l=0}^{K-1}C(l,K-1)p^{l}(1-p)^{K-1-l}
\int \prod_{k=1}^ldv_k\rho(v_k)e^{-\beta\ln(\frac{2\pi}{K-\sum_kv_k})}\delta(v-\frac{1}{K-\sum_kv_k}).
\end{align}
with $C(l,K)=K!/(l!(K-l)!)$. Again we solve these equations with a population dynamics algorithm as described in Appendix \ref{app:pop}. The only difference is that here we need only one population of variables $(c^a,v^a)$ and the Boltzmann weights $e^{-2\beta \Delta g_{\to}^a}$.

After solving the above equations for $p$ and $\rho(v)$, the free energy is given by
\begin{align}
-\beta f_g=\ln(z_0(K)+z_1(K))-\frac{K}{2}\ln(z_{00}+2z_{01}+z_{11}),
\end{align}
where $z_0(K), z_1(K)$ are as before and
\begin{align}
z_{00}&=(1-p)^2,\\
z_{01}&=(1-p)p,\\
z_{11}&=p^2\int dv dv'\rho(v)\rho(v')e^{\beta\ln(1-vv')}.
\end{align}
Recall that 
\begin{align}
-\beta f=\frac{1}{N}\ln Z=-\beta (f_g-\ln(2\pi)).
\end{align}

On the other hand, the average energy density is
\begin{align}
e=-\frac{1}{N}\langle\ln\mathrm{det}A(\mathbf{c})\rangle=2\langle g\rangle-\ln(2\pi).
\end{align}
Here
\begin{align}
2\langle g\rangle =\frac{g_0(K)+g_1(K)}{z_0(K)+z_1(K)}-\frac{K}{2}\frac{g_{11}}{z_{00}+2z_{01}+z_{11}},
\end{align}
with
\begin{align}
g_0(K)=z_0(K)\ln(2\pi),
\end{align}
\begin{align}
g_1(K) =\sum_{l=0}^{K}C(l,K)p^{l}(1-p)^{K-l}
\int \prod_{k=1}^ldv_k\rho(v_k)\ln(\frac{2\pi}{K-\sum_kv_k})e^{-\beta\ln(\frac{2\pi}{K-\sum_kv_k})},
\end{align}
\begin{align}
g_{11}=-p^2\int dv dv'\rho(v)\rho(v')\ln(1-vv')e^{\beta\ln(1-vv')}.
\end{align}
 
Finally, by a Legendre transformation, the entropy is obtained
\begin{align}
s=\beta(e-f).
\end{align}
A parametric plot of this quantity vs $e$ gives the entropy spectrum $s(e)$.

\section{Minors of homogeneous complete graphs: a mean-field model}\label{app:MF}
Consider a symmetric matrix $\mathbf{A}$ associated to a complete graph of size $N$ with elements $A_{ij}=\Lambda\delta_{ij}-\Gamma(1-\delta_{ij})$. We assume that $\Lambda>0$ and $\Lambda\ge N\Gamma >0$ to have a positive and diagonally dominant matrix. A minor configuration of size $l$ here is represented by a $l\times l$ matrix with diagonal elements $\Lambda$ and off-diagonal elements $-\Gamma$.
Such a matrix has one eigenvalue $\Lambda-(l-1)\Gamma$ and $l-1$ eigenvalues $\Lambda+\Gamma$. Thus the determinant (for $0<l\le N$) is given by
\begin{align}
\mathrm{det}(\mathbf{A}(\mathbf{c}))=[\Lambda-(l-1)\Gamma][\Lambda+\Gamma]^{l-1},
\end{align}
and the energy is
\begin{align}
E_l=-\ln(\mathrm{det}(\mathbf{A}(\mathbf{c})))=-(l-1)\ln(\Lambda+\Gamma)-\ln(\Lambda-(l-1)\Gamma).
\end{align}
The partition function is
\begin{align}
Z=\sum_{\mathbf{c}}e^{\beta \ln \mathrm{det}(\mathbf{A}(\mathbf{c}))}=1+\sum_{l=1}^{N-1}\Omega(l,N)e^{-\beta E_l},
\end{align}
where $\Omega(l,N)=N!/(l!(N-l)!)$. Or, in terms of the energy and entropy densities
\begin{align}
Z\simeq N\int_0^1 dx e^{N[s(x)-\beta e(x)]}=N\int_0^1 dx e^{-N\beta f(x)},\\
\end{align}
where $x=l/N$ is the number density of the present indices, and 
\begin{align}
s(x) &=-x\ln(x)-(1-x)\ln(1-x),\\
e(x) &=-x\ln(\Lambda+\Gamma)-\frac{1}{N}\ln(\Lambda-N\Gamma x),\\
f(x) &=e(x)-\frac{1}{\beta}s(x).
\end{align}
The thermodynamic behavior depends on the scaling of the matrix elements $\Lambda,\Gamma$ with $N$.

\subsection{Scaling limit I: $\Lambda=N-1, \Gamma=1$}
This is Laplacian of a complete graph. Here we use the scaling $\beta \to \beta/\ln(N)$ and replace $e(x)/\ln(N)$ with $e(x)$ as $N\to \infty$. Ignoring the sub-extensive terms in the energy density, we get 
\begin{align}
e(x) =-x.
\end{align}
The minimum of the free energy function 
\begin{align}
f(x) =e(x)-\frac{1}{\beta}s(x),
\end{align}
at $x^*$ determines the equilibrium values of the main quantities $e(x^*), s(x^*), f(x^*)$. Recall that $x^*$ is the presence probability $p$ we used in the main text.
Here the free energy is always minimized for a number density $\frac{1}{2}\le x^*<1$, which is the sole solution to 
\begin{align}
\beta=\ln(\frac{x^*}{1-x^*}). 
\end{align}
This system dose not display a phase transition. 
By increasing $\beta$ from zero to infinity, the density of relevant minors $x^*$ increases smoothly from $1/2$ to $1$. And, the minimum of free energy function remains an extremum ($\frac{df}{dx}(x^*)=0$) as $\beta\to \infty$.
There are $N$ ground state configurations of energy $E_{N-1}=-(N-2)\ln N$ where only one of the nodes is not present in the minor configuration. So, the Hamming distance between two ground states is $2$.

\subsection{Scaling limit II: $\Lambda=\mathrm{finite}, \Gamma=\gamma/N$}
This can be considered as a mean-field model of random regular graphs of degree $K=\Lambda$. Here the energy density as a function of $x$ in the thermodynamic limit reads
\begin{align}
e(x) =-x\ln(\Lambda).
\end{align}
Note that value of $\Gamma$ is not relevant in both the above scaling limits. Again, we do not observe a finite-temperature phase transition; the minimum of free energy at $x^*$ changes smoothly as $\beta$ approaches infinity. That is equation
\begin{align}
\frac{df}{dx}(x^*)=-\ln \Lambda+\frac{1}{\beta}\ln(\frac{x^*}{1-x^*})=0,
\end{align}
always has a unique solution $0<x^*<1$ due to the monotonic behaviors of the two parts of the equation
\begin{align}
\beta \ln \Lambda=\ln(\frac{x^*}{1-x^*}).
\end{align}
However, depending on the value of $\Lambda$ the solution $x^*$ approaches $0$ or $1$ as $\beta\to \infty$ for $\Lambda<1$ and $\Lambda>1$, respectively. For $\Lambda=1$ the solution is always $x^*=\frac{1}{2}$. The ground states have a trivial structure: When $\Lambda<1$ the minimum energy is obtained by $N$ minor configurations with a single present index, putting aside the all-zero configuration. When $\Lambda>1$ the minimum energy is obtained by $N$ minor configurations with a single absent index. The case $\Lambda=1$ has the maximal ground state degeneracy where all minor configurations are equivalent in energy. Note that in all the above scenarios the value of entropy at zero-temperature coincides with the limit $\beta\to \infty$ of the entropy function. It means that the entropy changes continuously as the temperature increases.

\end{document}